\documentclass[a4paper,10pt]{article}
\usepackage[affil-it]{authblk}
\usepackage{bm}
\usepackage{multicol, caption}
\usepackage{graphicx}
\usepackage{titlesec}
\usepackage[T1]{fontenc}

\renewcommand{\thesection}{\Roman{section}.}
\renewcommand{\thesubsection}{\Alph{subsection}.}
\renewcommand{\thesubsubsection}{\arabic{subsubsection}.}

\newenvironment{Figure}
  {\par\medskip\noindent\minipage{\linewidth}}
  {\endminipage\par\medskip}

\titleformat{\section}{\centering\small\scshape\bfseries}{\thesection}{1 em}{}
\titleformat{\subsection}{\centering\small\bfseries}{\thesubsection}{1em}{}
\titleformat{\subsubsection}{\centering\small\em}{\thesubsubsection}{1em}{}

\title{\large \bf{Flux-cutting and flux-transport effects in type-II superconductor slabs in a parallel rotating magnetic field}}

\author[1]{\small{R. Cort\'es-Maldonado}}
\affil[1]{\small{Instituto de F\'isica, Benem\'erita Universidad Aut\'onoma de Puebla,

Apdo. Post. J-48, Puebla, Pue. 72570, Mexico}}

\author[2]{\small{J.E. Espinosa-Rosales}}
 \affil[2]{\small{Facultad de Ciencias F\'{\i}sico-Matem\'aticas, Benem\'erita Universidad Aut\'onoma de Puebla,

Apdo. Post. 1152, Puebla, Pue., 72000, Mexico}}

\author[3]{\small{A.F. Carballo-S\'anchez}}
\affil[3]{\small{Universidad del Istmo, Campus Tehuantepec, Tehuantepec, Oax., 70760, Mexico}}

\author[1]{\small{F. P\'erez-Rodr\'iguez \footnote{Author to whom correspondence should be addressed: fperez@ifuap.buap.mx}}}
\date{}
\begin{document}
\maketitle
\begin{abstract}
The magnetic response of irreversible type-II superconductor slabs
subjected to in-plane rotating magnetic field is investigated by
applying the circular, elliptic, extended-elliptic, and rectangular
flux-line-cutting critical-state models. Specifically, the models
have been applied to explain experiments on a PbBi rotating disk in
a fixed magnetic field ${\bm H}_a$, parallel to the flat surfaces.
Here, we have exploited the equivalency of the experimental
situation with that of a fixed disk under the action of a parallel
magnetic field, rotating in the opposite sense.  The effect of both
the magnitude $H_a$ of the applied magnetic field and its angle of
rotation $\alpha_s $ upon the magnetization of the superconductor
sample is analyzed. When $H_a$ is smaller than the penetration field
$H_P$, the magnetization components, parallel and perpendicular to
${\bm H_a}$, oscillate with increasing the rotation angle. On the
other hand, if the magnitude of the applied field, $H_a$, is larger
than $H_P$, both magnetization components become constant functions
of $\alpha_s$ at large rotation angles. The evolution of the
magnetic induction profiles inside the superconductor is also
studied.

\end{abstract}

\indent \indent PACS numbers: 74.25.Ha, 74.25.Op, 74.25.Sv, 74.25.Wx
\\
\indent \indent Keywords: flux cutting, flux transport, vortex pinning, critical state, hard superconductor


\begin{multicols}{2}
\section{Introduction}

The discovery of the phenomenon known as {\em quasisymmetrical
collapse} of magnetization \cite{Fis96}, which is observed in
superconductors subjected to crossed magnetic fields and well
interpreted within the simple Bean's critical-state model
\cite{Bea62,Bea70}, has been a turning point in the understanding of
the magnetic behavior of hard (irreversible type-II)
superconductors. Until then, the generalized double critical-state
model (GDCSM) \cite{Cle82,Cle84,Per85a,Per85b,Per85c}, which is
based on fundamental physical concepts such as flux transport and
flux-line-cutting \cite{Wal72,Cam72}, was successfully employed to
explain a variety of experiments where flux cutting occurs
\cite{Cav82,Boy77,Boy80,LeB84,Per97,Sil98}. An important feature of
the GDCSM is the assumption that flux cutting and flux depinning do
not affect each other. Besides, the GDCSM is inherently anisotropic
because the thresholds for these two effects are given by two
independent parameters, namely the critical current densities
parallel ($J_{c\parallel}$) and perpendicular ($J_{c\perp}$) to the
local magnetic induction ${\bm B}$. However, since the GDCSM cannot
reproduce the features of magnetic moment {\em collapse}
\cite{Fis00,Vol10}, whereas isotropic Bean's model does it, the main
assumption of the GDCSM has been questioned, motivating the
development of new critical-state models in the past few years.

In Ref. \cite{Rom03c},  the so-called elliptic flux-line-cutting
critical-state model was proposed.  This model introduces the
anisotropy, induced by flux-line-cutting effects, by using a
procedure similar to that for structurally anisotropic
superconductors \cite{Vol01,Rom03b}, i.e. the magnitude of the
critical current density $J_c$, being the only parameter used within
the isotropic Bean's model, is substituted by a symmetrical tensor
$(J_c)_{ik}$ with principal values $J_{c\parallel}$ and
$J_{c\perp}$, corresponding to the directions along and  across the
local magnetic induction ${\bm B}$. In good agreement with the
experiment on YBa$_2$Cu$_3$O$_{7-\delta}$ samples
\cite{Fis96,Fis00}, the elliptic critical-state model predicts the
quasisymmetrical suppression of the average magnetization $<M_z>$,
for paramagnetic and diamagnetic initial states, by sweeping a
transverse field $H_y$ of magnitude much smaller than dc-bias
magnetic field $H_z$ \cite{Rom03c,Rom04}. When the magnitudes of the
crossed fields $H_y$ and $H_z$ are comparable, the value of the
magnetization $<M_z>$ after many cycles of the transverse field
$H_y$ turns out to be positive for both diamagnetic and paramagnetic
initial states if $J_{c\parallel}> J_{c\perp}$. To our knowledge,
the observation of such a {\em paramagnetism } of hard
superconductors was first reported in Refs. \cite{Fis97a,Fis97b}.
The elliptic model also describes the behavior of $<M_y>(H_y)$  and
$<M_z>(H_y)$ in crossed fields $H_y$ and $H_z$ \cite{Rom03c,Rom05},
which was observed in the experiments  on a VTi ribbon with {\em
nonmagnetic} initial state \cite{LeB84,Lor79}. Here, the good
agreement with the experiment was achieved by using a relatively
large anisotropy parameter $J_{c\parallel}/J_{c\perp}=6$. It should
be noticed that the Bean's critical state model predicts neither the
phenomenon of the {\em paramagnetism} of hard superconductors nor
the behavior of the components of the average magnetization found in
Refs. \cite{LeB84,Lor79}. Furthermore, as it is shown in Refs.
\cite{Rom03c,Rom08}, the elliptic critical-state model successfully
describes the magnetic response of superconducting disks undergoing
oscillations in a magnetic field of fixed magnitude for nonmagnetic,
paramagnetic, and diamagnetic initial states \cite{Cav82}.

Despite the great success of the elliptic model \cite{Rom03c}, it
turns out that there exist phenomena, associated with flux cutting,
which are not completely described within such a model. So, in a
very recent work \cite{Cle11b}, the elliptic critical-state model
and other four theoretical approaches for describing the critical
state of type-II superconductors (GDCSM, extended GDCSM
\cite{Bra07,Mik10}, extended elliptic critical-state model
\cite{Cle11b,Cle11a}, and an elliptic critical-state model based on
the variational principle \cite{Bad09}) were tested. There, the
angular dependencies of the critical current density $J_c$ and the
electric field ${\bm E}$ (for $J$ just above $J_c$) were measured,
using an epitaxially grown YBCO thin film, and compared with the
predictions of the five theories. The measurements of angular
dependence of the critical-current density $J_c$ demonstrated a
behavior rather similar to that assumed by the elliptic
critical-state models. Besides, the smooth angular dependence of the
ratio of the transverse to the longitudinal components of the
electric field $E_y/E_z$  for $J$ just above $J_c$, predicted by the
three elliptic models, was verified in the experiment \cite{Cle11b}.
However, the original critical-state model \cite{Rom03c} leads to
small values of the ratio $E_y/E_z$ in comparison with the
experimental data and the results obtained from the other two
elliptic models. On the basis of this detailed comparison between
experiment and the five theories, it was concluded in Ref.
\cite{Cle11b} that the experiment favors only one of the models,
namely the extended elliptic critical-state model.

The aim of the present work is to investigate the behavior of a hard
superconductor in a parallel rotating magnetic field (or
equivalently, the response of a rotating superconductor in a fixed
magnetic field) and to compare the predictions of four
critical-state models with experiment.  Concretely, we shall
consider the Bean's critical-state model \cite{Bea62,Bea70}, the
original elliptic critical-state model \cite{Rom03c,Rom04}, the
recently-proposed extended elliptic model \cite{Cle11b,Cle11a}, as
well as the GDCSM \cite{Cle82,Cle84,Per85a,Per85b,Per85c}, whose
main characteristics and assumptions will be revisited in Sec.
\ref{Theory}. We shall numerically solve Maxwell equations with the
material equation postulated by each of the considered
critical-state models to calculate magnetization curves for a
superconductor disk rotating in a fixed magnetic field as in the
experiment \cite{Sek89} (Sec. \ref{Comparison}). Here, we shall
analyze the effect of the magnitude $H_a$ of the applied magnetic
field upon the dependencies of the magnetization components,
parallel and perpendicular to ${\bm H}_a$, on the rotation angle of
the superconductor disk. The evolution of magnetic induction
profiles will also be studied to explain the magnetic response of
the rotating hard-superconductor sample.

\section{Theoretical formalism}\label{Theory}
 Let us consider a superconducting slab of thickness $d$, which occupies the space
 $0<x<d$ and is subjected to a magnetic field ${\bm H}_a$ parallel  to its
 surfaces:
 \begin{equation}\label{Ha}
{\bm H}_a=H_a\hat{{\bm \alpha}_s}=H_a[\hat{{\bm
y}}\sin(\alpha_s)+\hat{{\bm z}}\cos(\alpha_s)],
 \end{equation}
where $\alpha_s$ is the angle of the applied magnetic field ${\bm
H}_a$ with respect to the $z$-axis. Hence, the magnetic induction
${\bm B}(x,t)$ inside the superconducting slab can be expressed as
\begin{equation}\label{B}
{\bm B}=B(x,t)[\hat{{\bm y}}\sin(\alpha(x,t))+\hat{{\bm
z}}\cos(\alpha(x,t))],
\end{equation}
where $B$ and $\alpha$ are respectively the magnitude and the tilt
angle of the magnetic induction. It is convenient to write the
electric field ${\bm E}(x,t)$ and the electrical current density
${\bm J}(x,t)$ in terms of their components parallel and
perpendicular to the local magnetic induction ${\bm B}(x,t)$:
\begin{eqnarray}\label{EB}
{\bm E}(x,t)&=&E_{\parallel}(x,t) \hat {{\bm
\alpha}}(x,t)+E_{\perp}(x,t) \hat {{\bm \beta}}(x,t), \\
{\bm J}(x,t)&=&J_{\parallel}(x,t) \hat {{\bm
\alpha}}(x,t)+J_{\perp}(x,t) \hat {{\bm \beta}}(x,t), \label{J}
\end{eqnarray}
where $\hat{{\bm \beta}}(x,t)= \hat{{\bm x}}\times \hat{{\bm \alpha
}}(x,t)$. Inside the superconductor sample, we shall assume that the
magnetic induction and the magnetic field satisfy the relation ${\bm
B}(x,t)=\mu_0{\bm H}(x,t)$, which is good enough for applied
magnetic fields much larger than the first critical field ($H_a\gg
H_{c1}$). Moreover, any surface barrier against the flux entry (or
exit) will be neglected. According to the planar geometry of the
problem, we can rewrite Ampere and Lorentz laws,
\begin{eqnarray}
\nabla \times {\bm B}(x,t)& =& \mu_0 {\bm J}(x,t), \\
 \nabla \times
{\bm E}(x,t)&=& -\frac{\partial {\bm B}}{\partial t},
\end{eqnarray}
as follow
\begin{eqnarray}
\frac{\partial B}{\partial x}&=& - \mu_0 J_{\perp},  \label{ME1} \\
B\frac{\partial \alpha}{\partial x}&=& -\mu_0 J_{\parallel},
\label{ME2}
\end{eqnarray}
\begin{eqnarray}
\frac{\partial E_{\perp}}{\partial x}+E_{\parallel}\frac{\partial
\alpha}{\partial x}&=&-\frac{\partial B}{\partial t}, \label{ME3}\\
E_{\perp}\frac{\partial \alpha}{\partial x}-\frac{\partial
E_{\parallel}}{\partial x}&=&-B\frac{\partial \alpha}{\partial t}.
\label{ME4}
\end{eqnarray}
To solve the resulting system of differential equations for ${\bm
E}$, ${\bm B}$ and ${\bm J}$, one should add the material equation.
Below, we shall use the material equations corresponding to the
 circular, elliptic, extended-elliptic, and rectangular
 flux-line-cutting critical-state models.

\subsection{Circular model}

The first model for describing the magnetic behavior of
superconductors in multicomponent situations  was proposed by Bean
\cite{Bea62,Bea70}. According to it, the critical current density
${\bm J}$ points always along the local electric field ${\bm E}$.
Hence,
\begin{equation}\label{Bm}
{\bm J}=J_c\frac{{\bm E}}{E}.
\end{equation}
The magnitude of the critical current density $J=J_c$ is the unique
phenomenological parameter used and may depend on the magnitude of
the magnetic induction $B$. In the planar geometry [see
Eqs.~(\ref{Ha})-(\ref{J})], the assumption $J=J_c$ corresponds to a
circle in the $J_{\perp}$-$J_{\parallel}$ plane.

In numerically solving the system of equations
(\ref{ME1})-(\ref{ME4}) for the electromagnetic fields, it is
necessary to rewrite Eq. (\ref{Bm}) as
\begin{equation}
{\bm E}=E(J)\frac{{\bm J}}{J}, \label{Bm2a}
\end{equation}
\begin{equation}
 E(J) = \left\{
\begin{array}{ll}
0,   & J \leq J_{c}(B) \\
\rho (J-J_{c}(B)), & J \geq J_{c}(B)
\end{array}
     \right.
 \label{Bm2b}
\end{equation}
where $\rho$ is an effective resistivity. It should be mentioned
that for slow variations of the surface boundary conditions,
producing a small magnitude of the induced electric field ($E\ll
\rho J_c$), the magnetic induction profiles are practically relaxed
and independent of the parameter $\rho $ \cite{Rom03a}.

\subsection{Elliptic model}

The elliptic flux-line cutting critical-state model
\cite{Rom03c,Rom04,Rom05} postulates:
\begin{equation}\label{ellm}
J_i=(J_c)_{ik} \frac{E_k}{E},
\end{equation}
where
\begin{equation}
(J_c)_{ik}=J_{c,i}(B) \, \delta_{ij}, \qquad i,k=\perp , \parallel.
\end{equation}
Here $\delta _{ik}$ is the Kronecker delta symbol. Within the
elliptic critical-state model (\ref{ellm}),  the magnitude of the
critical current density $J_c$ draws an ellipse on the
$J_{\perp}$-$J_{\parallel}$ plane. This model makes use of two
phenomenological parameters, namely the extreme values $J_{c\perp}$
and $J_{c\parallel}$ for the radius of the ellipse drawn by the
magnitude of the critical current density. In the numerical
calculations for solving the system of equations
(\ref{ME1})-(\ref{ME4}), the relation (\ref{ellm}) is rewritten in
the form
\begin{equation}\label{elm}
E_i=E(J)\left ( J_c^{-1}  \right )_{ik} J_k,
\end{equation}
\begin{equation}
 E(J) = \left\{
\begin{array}{ll}
0,   & J \leq J_{c}(B,\phi) \\
\rho (J-J_{c}(B,\phi)), & J \geq J_{c}(B,\phi)
\end{array}
     \right. ,
 \label{elm2}
\end{equation}
where $\left (J_c^{-1}\right )_{ik}  $ is the inverse of the matrix
$(J_c)_{ik}$ in (\ref{ellm}). The magnitude of the critical current
density, $J_{c}(B,\phi )$, is given by the expression
\begin{equation}\label{Jcphi}
J_{c}(B,\phi ) =\left [\frac{\cos^2 (\phi)}{J_{c\parallel}^2(B)}
+\frac{\sin^2 (\phi)}{J_{c\perp}^2(B)} \right ]^{-1/2}.
\end{equation}
Here, $\phi $ denotes the angle of the critical current density
${\bm J}$ with respect to the direction of the flux density ${\bf
B}$. If $J_{c\perp}=J_{c\parallel}$, the elliptic critical-state
model (\ref{elm}) goes over into the Bean's (circular)
critical-state model (\ref{Bm2a}). Besides, the calculations of
electromagnetic fields with $J$ close to $J_c$ are also independent
of the auxiliary parameter $\rho $ in Eq. (\ref{elm2}).

\subsection{Extended elliptic model}

The elliptic critical-state model, described in previous subsection,
has recently been extended in Refs. \cite{Cle11b,Cle11a} by
introducing the general relations
\begin{eqnarray}
E_{\perp}&=&\rho_{\perp} J_{\perp},\\
E_{\parallel}&=&\rho_{\parallel} J_{\parallel},
\end{eqnarray}
where $\rho_{\perp}$ and $\rho_{\parallel}$ are nonlinear effective
resistivities, having a ratio $r=\rho_{\parallel}/\rho_{\perp}$
independent of $J$ just above $J_c$ as it was experimentally found
\cite{Cle11b}. A model for the effective resistivities is given by
\cite{Cle11a}
\begin{equation}\label{eel1}
 E_{\perp} = \left\{
\begin{array}{ll}
0,   & 0\leq \mid J_{\perp} \mid \leq J_{cd} \\
\rho_{d} (\mid J_{\perp}\mid -J_{cd}){\rm sign}(J_{\perp}), & \mid
J_{\perp}\mid \geq J_{cd}
\end{array}
     \right. ,
\end{equation}
\begin{equation}\label{eel2}
 E_{\parallel}= \left\{
\begin{array}{ll}
0,   & 0\leq \mid J_{\parallel} \mid \leq J_{cc} \\
\rho_{c} (\mid J_{\parallel}\mid -J_{cc}){\rm sign}(J_{\parallel}),
& \mid J_{\parallel} \mid \geq J_{cc}
\end{array}
     \right. .
\end{equation}
Here, the subscripts ``d" and ``c" respectively refer to depinning
and cutting. Besides, $J_{cd}= J_{c}(B,\phi) \mid \sin (\phi)\mid$
and $J_{cc}=J_{c}(B,\phi) \mid \cos (\phi)\mid$, where
$J_{c}(B,\phi)$ is defined according to the elliptic critical-state
model as in Eq. (\ref{Jcphi}). If $\mid J-J_c \mid/J_c\ll 1$, the
extended elliptic critical-state model reduces to the original one
[Eqs (\ref{elm}) and (\ref{elm2})] by replacing $\rho_d$ and
$\rho_c$  in Eqs. (\ref{eel1}) and (\ref{eel2}) with $\rho
J_c/J_{c\perp}$ and $\rho J_c/J_{c\parallel}$, correspondingly.
Hence, in the case of the original elliptic model, the ratio
$r=\rho_{\parallel}/\rho_{\perp}$ at $J>J_c$ is equal to
$J_{c\perp}/J_{c\parallel}$. On the other hand, the extended
elliptic critical-state model is capable to modify the relation
between the components of the electric field ${\bm E}$ and the
current density ${\bm J}$ with the aid of the additional parameter
$r$.

\subsection{Rectangular model}
The generalized double critical-state model (GDCSM)
\cite{Cle82,Cle84,Per85a,Per85b,Per85c} uses two phenomenological
parameters, namely the critical values, $J_{c\parallel}$ and
$J_{c\perp}$, of the electrical current density along and
perpendicular to the local magnetic induction. Within this model,
each component of the electrical current density is determined by
its own electric field as
\begin{eqnarray}\label{GDCSM}
J_{\perp}=J_{c\perp}\, {\rm sign}(E_{\perp}), \\
J_{\parallel}=J_{c\parallel} \, {\rm sign}(E_{\parallel}).
\end{eqnarray}
Evidently, the magnitude of the critical current density traces a
rectangle in the $J_{\perp}$-$J_{\parallel}$ plane. The parameter
$J_{c\perp}$ determines the threshold for depinning of vortices,
whereas $J_{c\parallel}$ indicates the onset of flux-line cutting in
the vortex array. In calculating the electromagnetic fields within
the GDCSM, the material equation (\ref{GDCSM}) is written in the
form
\begin{equation}\label{GDCSM1}
 E_{\perp} = \left\{
\begin{array}{ll}
0,   & 0\leq \mid J_{\perp} \mid \leq J_{c\perp} \\
\rho_{\perp} (\mid J_{\perp}\mid -J_{c\perp}){\rm sign}(J_{\perp}),
& \mid J_{\perp}\mid \geq J_{c\perp}
\end{array}
     \right. ,
\end{equation}
\begin{equation}\label{GDCSM2}
 E_{\parallel} = \left\{
\begin{array}{ll}
0,   & 0\leq \mid J_{\parallel} \mid \leq J_{c\parallel} \\
\rho_{\parallel} (\mid J_{\parallel}\mid -J_{c\parallel}){\rm
sign}(J_{\parallel}), & \mid J_{\parallel} \mid \geq J_{c\parallel}
\end{array}
     \right. .
\end{equation}
The quantities  $\rho_{\perp}$ and $\rho_{\parallel}$ are effective
flux-flow and flux-line-cutting resistivities of the material.
However, unlike the above-commented critical-state models, the GDCSM
allows the existence of zones in the $J_{\perp}$-$J_{\parallel}$
plane where either  flux cutting or flux transport exclusively
occur. The latter is possible due to the assumption of the GDCSM
that the threshold for flux depinning, $J_{c\perp}$ (flux cutting,
$J_{c\parallel}$) is independent of the component $J_{\parallel}$
($J_{\perp}$) [compare Eqs. (\ref{GDCSM1}) and (\ref{GDCSM2}) with
Eqs. (\ref{eel1}) and (\ref{eel2}) where $J_{cd}$ and $J_{cc}$
depend on the angle $\phi=\arctan(J_{\perp}/J_{\parallel})$].

\section{Numerical results and comparison with
experiment}\label{Comparison}

In the present section we will apply the flux-line-cutting
critical-state models, commented above, to explain experimental
magnetization curves \cite{Sek89} of a PbBi superconducting disk,
rotating in the presence of an external magnetic field ${\bm H}_a$,
which is oriented parallel to the disk plane (along the $z$-axis)
and perpendicular to the axis of rotation.

\subsection{Experimental results}

Fig. \ref{F1}a exhibits a standard magnetization curve, which was
measured in Ref. \cite{Sek89}, for a PbBi disk of thickness
$d=0.8$mm. The hysteresis in Fig.~\ref{F1}a clearly corresponds to
the magnetization curve of a type-II irreversible superconductor
since its return crosses over and remains in the paramagnetic region
as a result of the strong flux pinning. In the experiment, the
isotropy of the PbBi disk was also verified by comparing standard
magnetization curves with ${\bm H}_a$ directed along different
diameters of the disk.

Panels (a)-(c) in Fig. \ref{F2} show graphs of the magnetization
components, $<M_y>=<B_y>/\mu_0$  and $-<M_z>= H_a-<B_z>/\mu_0$,
versus the angle $\theta$ of rotation,  measured in the work
\cite{Sek89} for the PbBi disk, rotating in the magnetic field ${\bm
H}_a$. The measurements started in the nonmagnetic initial state
which is reached after cooling the superconductor at the fields
$H_a/H_P=$ 0.5 (panel a), 1.0 (panel b), and 2.0 (panel c), where
 $H_P$ ($\mu_0 H_P=0.1015$T \cite{Sek89}) is the penetration
field. The initial state is supposed to be nonmagnetic because no
Meissner effect (flux expulsion) was observed after field cooling,
within the accuracy ($\Delta <M>\leq 1$ Gauss) of the experiment.

\begin{Figure}
\includegraphics[scale=0.8]{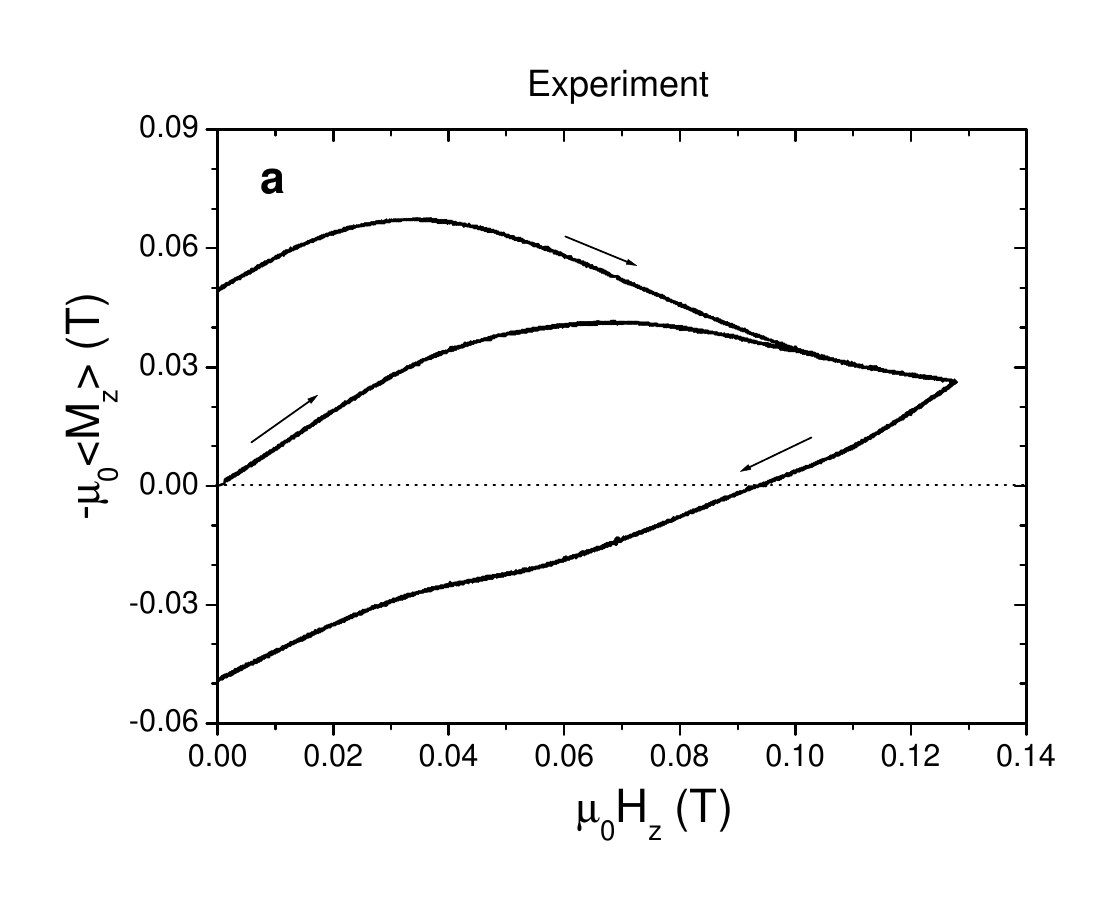}
\\
\includegraphics[scale=0.8]{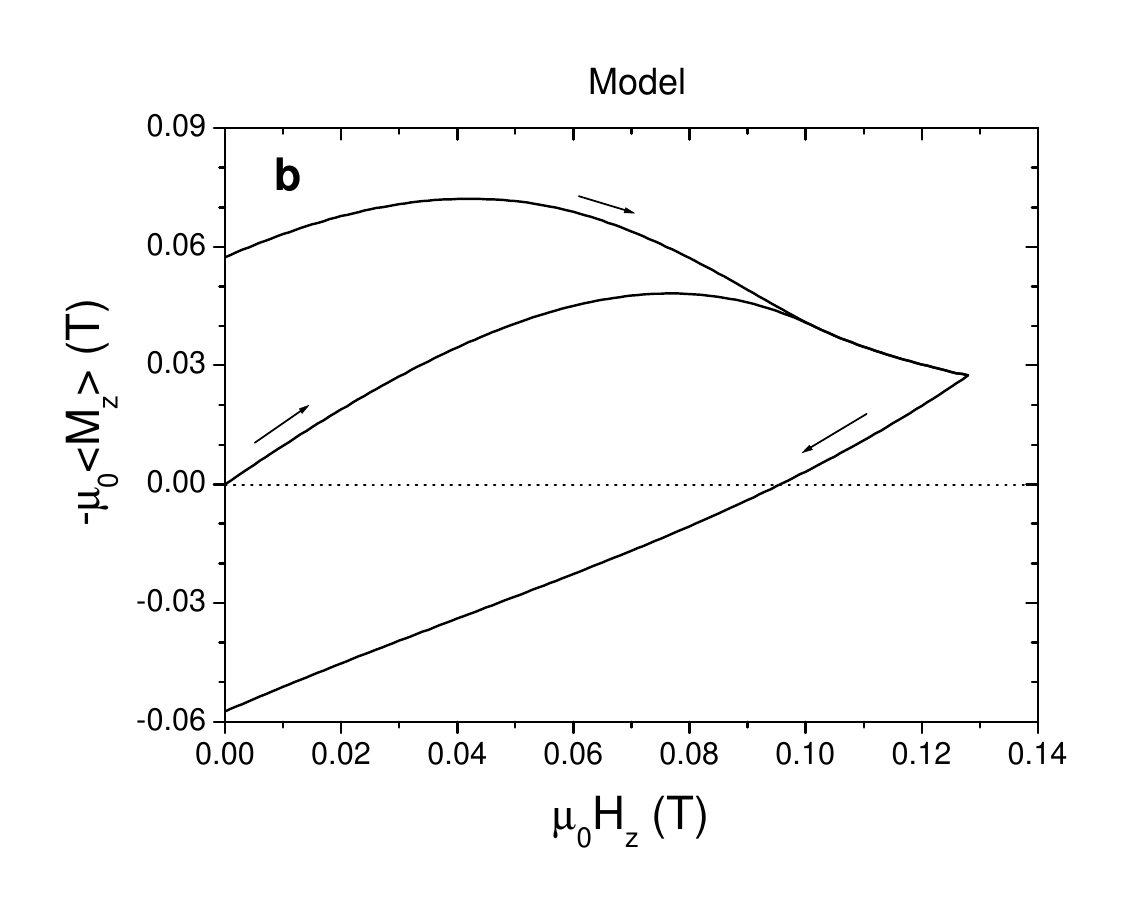}
\captionof{figure}{Standard magnetization curves (a) for a PbBi disk, taken
from Ref. \cite{Sek89}. Theoretical magnetization curves (b)
obtained with a critical current density $J_{c\perp}(B)$ as in Eq.
(\ref{jcperp}). \label{F1}}
\end{Figure}

 As it is seen in Fig. \ref{F2}, for the smallest
value of $H_a$ ($=0.5 H_P$, panel a), both magnetization components
have a nonmonotonic behavior as functions of $\theta$. Such a
behavior of magnetization has also been observed in Ref.
\cite{Cav82} during the initial rotation of a Nb disk undergoing
slow oscillations in a parallel field. The dependence of the
magnetization on $\theta$ radically changes at larger values of
$H_a$. So (see Fig. \ref{F2},b), at $H_a=H_P$ the functions
$<M_y>(\theta)$ and $-<M_z>(\theta )$ initially grow with $\theta$
and later (at $\theta
> 150^{\circ}$) they practically become  constants with close values
($<M_y>\approx -<M_z>$). Also note that $M_y$ has a maximum at
$\theta \approx 75^{\circ}$. For $H_a$ larger than the penetration
field $H_P$ (panel c), the function $-<M_z>(\theta )$ takes values
smaller than those for $<M_y>(\theta)$. Both of them are almost
constant functions, except at small rotation \ angles because \ of

\begin{Figure}
\includegraphics[scale=0.8]{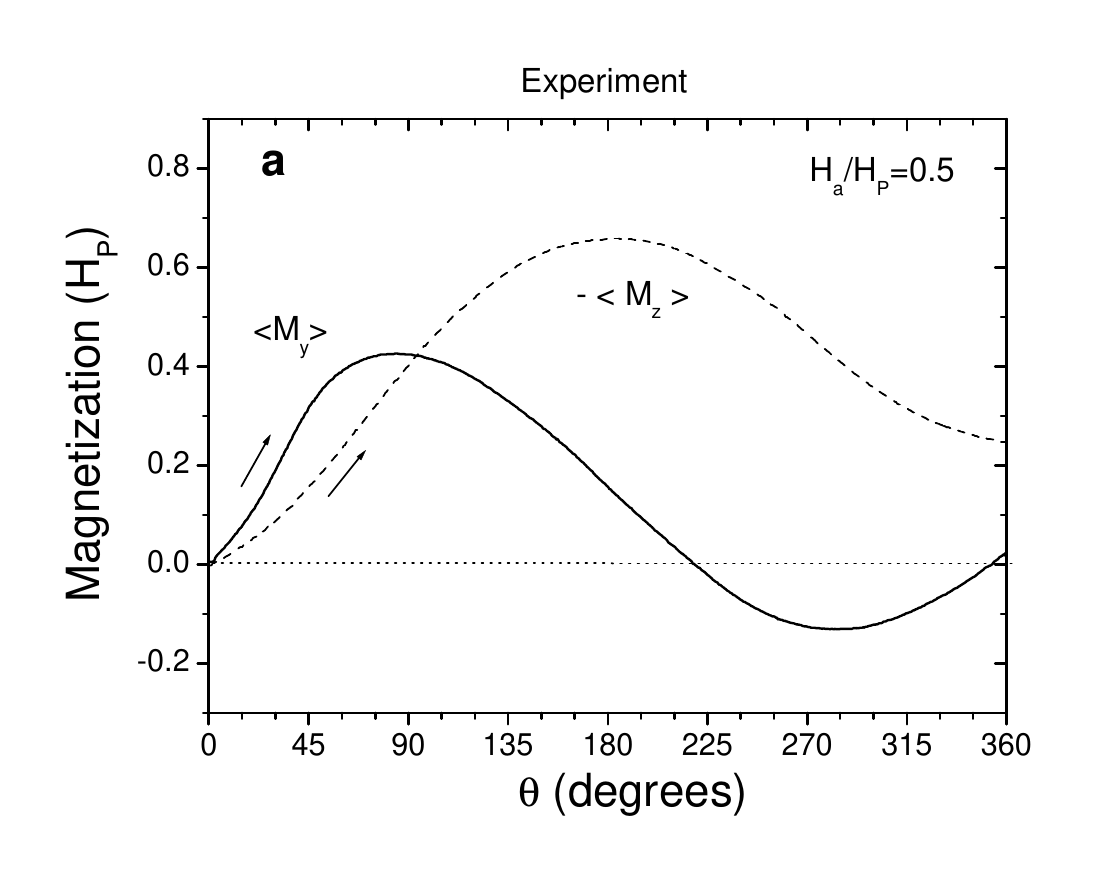}
\vspace{7mm}
\includegraphics[scale=0.8]{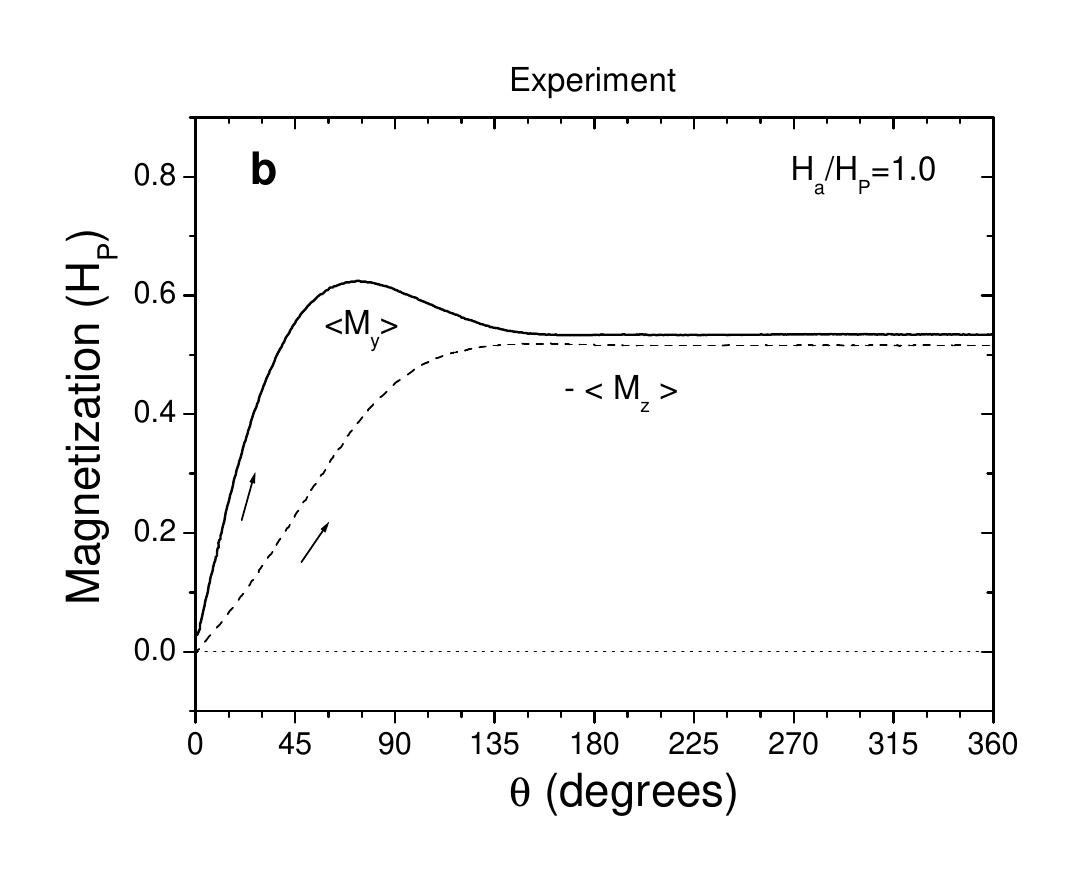}
\vspace{10mm}
\includegraphics[scale=0.8]{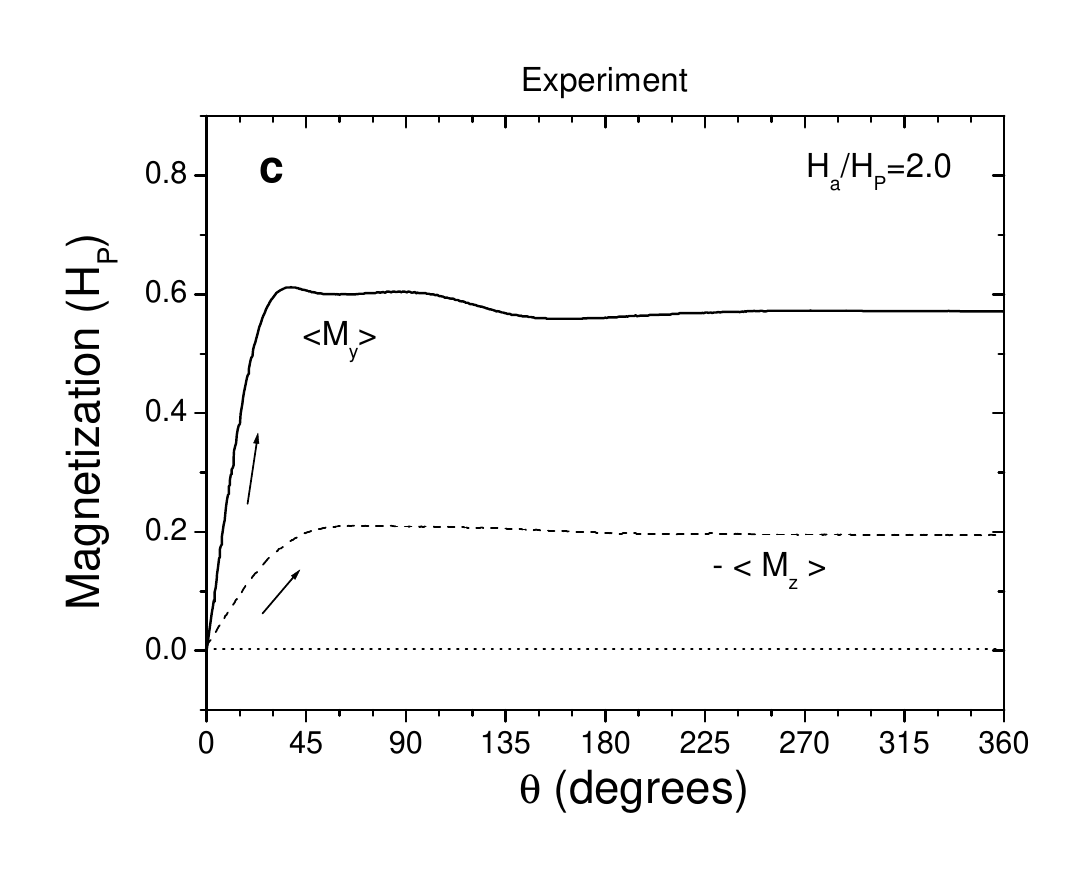}
\captionof{figure}{PbBi rotational curves measured in Ref.
\cite{Sek89}.\label{F2}}
\end{Figure}

\vspace{5mm}
\noindent
their fast initial growth. Thus, the maximum of $<M_y>$ is shifted to a
smaller value of $\theta$ ($ \approx 40^{\circ}$).

\subsection{Theoretical predictions}

The models described in the previous section can be applied to
explain the experimental results (Fig. \ref{F2}) if we fix the
sample and rotate the external magnetic field ${\bm H}_a$ (\ref{Ha})
by an angle $\alpha_s=-\theta$ instead of fixing the magnetic field
and rotating the superconducting sample. Then, the experimental
values $<M_y>$ and $-<M_z>$ should respectively correspond to the
quantities:
\begin{equation}\label{My}
<M_y>=  \frac{1}{\mu_0 d}\int_0^d dx B'_y(x),
\end{equation}
\begin{equation}\label{Mz}
-<M_z>=H_a-\frac{1}{\mu_0d}\int_0^d dx B'_z(x),
\end{equation}
where
\begin{equation}\label{byp}
B'_y=\hat{{\bm \alpha}}_s\times \hat{{\bm x}}\cdot {\bm
B}=B(x)\sin[\alpha(x)-\alpha_s],
\end{equation}
\begin{equation}\label{bzp}
B'_z=\hat{{\bm \alpha}_s}\cdot {\bm B}=B(x)\cos[\alpha(x)-\alpha_s].
\end{equation}

The calculations of magnetization components $<M_y>$ and $-<M_z>$
with the critical-state models, discussed in Sec. \ref{Theory},
 require the
 employment of the parameters $J_{c\perp}(B)$ and $J_{c\parallel}(B)$, depending on
 the magnetic induction. The
former, $J_{c\perp}(B)$,  is determined from the experimental curves
of magnetization versus the applied field, varying along one
direction only as in Fig.~\ref{F1} (In this case, flux cutting does
not occur and, consequently, the depinning effects are completely
responsible for the magnetic response of the superconductor). The
standard magnetization curves are well reproduced by any one of the
critical-state models (see above) with
\begin{equation}\label{jcperp}
J_{c\perp}(B)=\frac{J_{c\perp}(0)}{(1+B/\mu_0H_P)^{n_{\perp}}},
\end{equation}
$J_{c\perp}(0)= 47.11 \times 10^7$ A/m$^2$, and $n_{\perp}=2$
(compare panels (a) and (b) of Fig.~\ref{F1}). Other parameters of
the critical state models are found by adjusting theoretical
magnetization curves to the experimental ones (Fig. \ref{F2}).

\subsubsection{Circular model}

Within the Bean's circular critical-state model (\ref{Bm}), there is
only one phenomenological parameter, i.e.
$J_c(B)=J_{c\perp}(B)=J_{c\parallel}(B)$. Then, $J_c(B)$ has the
form (\ref{jcperp}) with the same values for the parameters
$J_{c\perp}(0)$, and $n_{\perp}$.

Fig. \ref{F3} shows our numerical results for $<M_y>$ and $-<M_z>$,
obtained with the Bean critical-state model. At first glance, it
seems that the circular model qualitatively reproduces  the
experimental magnetization curves (Fig. \ref{F2}). However, there
are important differences between its predictions and the
experiment.


\begin{Figure}
\includegraphics[scale=0.8]{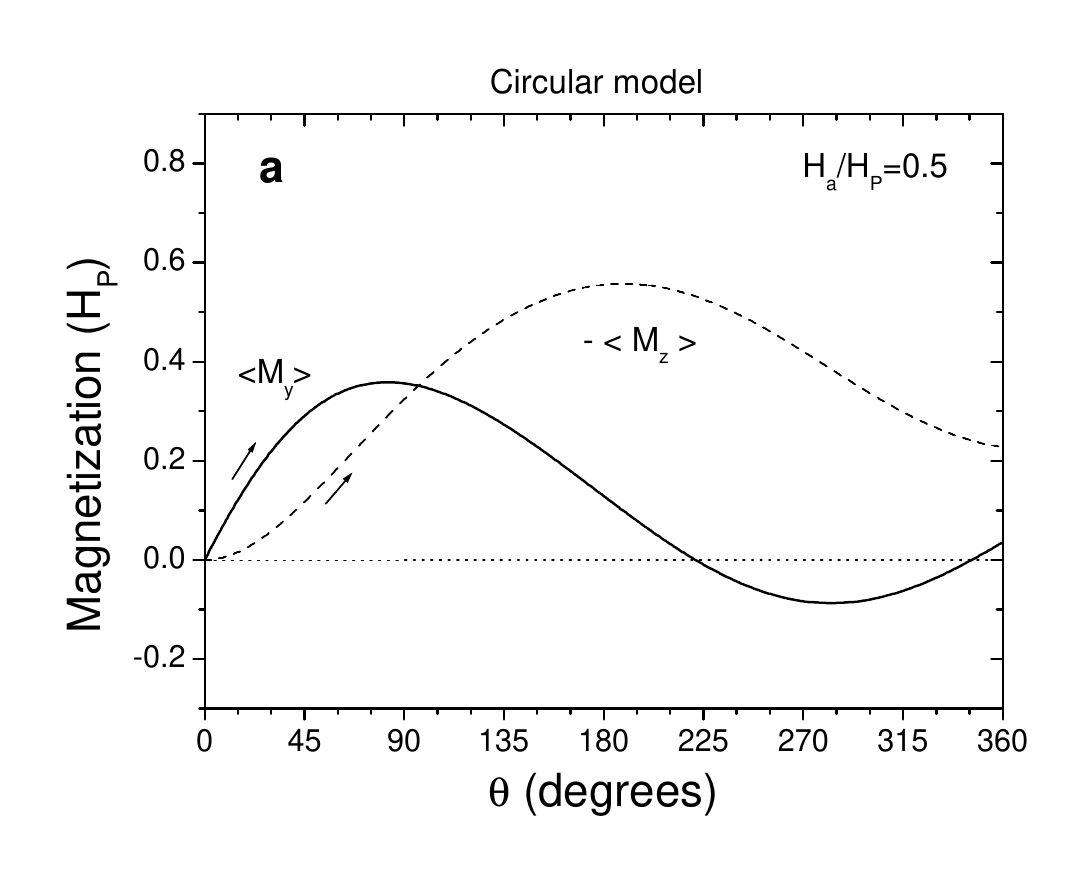}
\vspace{5mm}
\includegraphics[scale=0.8]{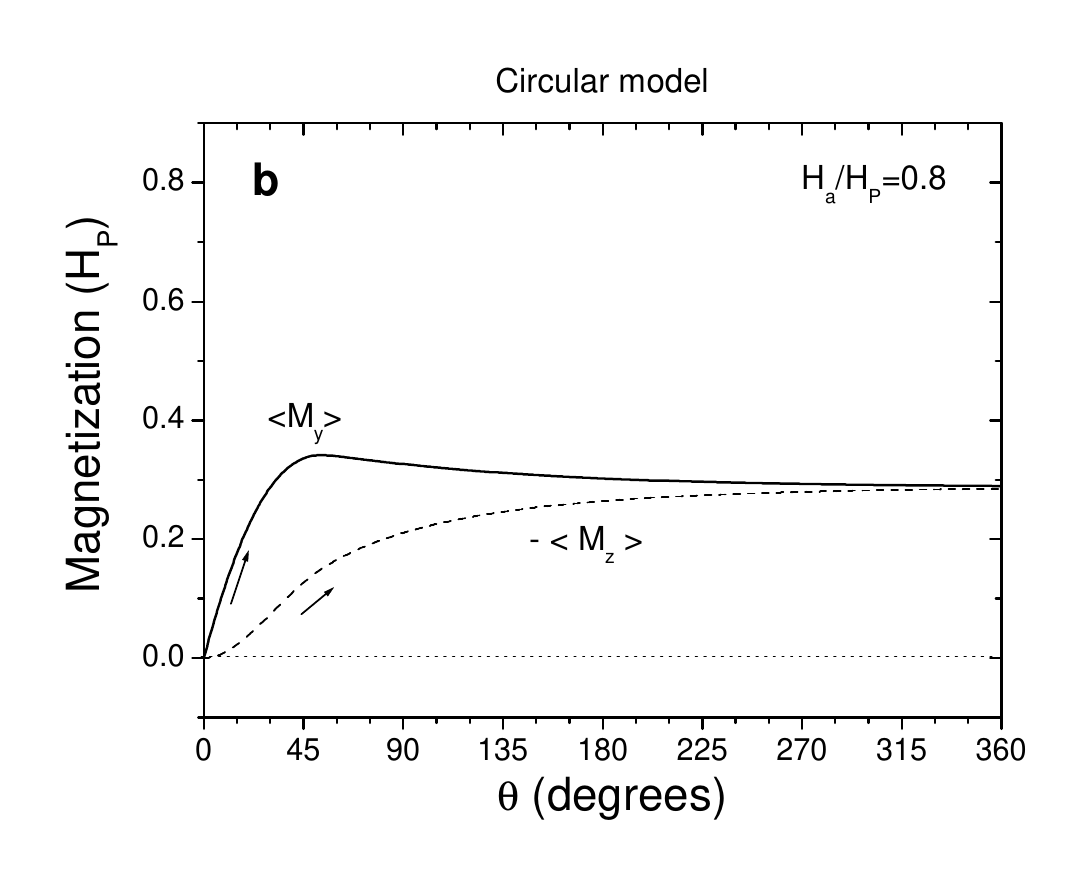}
\vspace{5mm}
\includegraphics[scale=0.8]{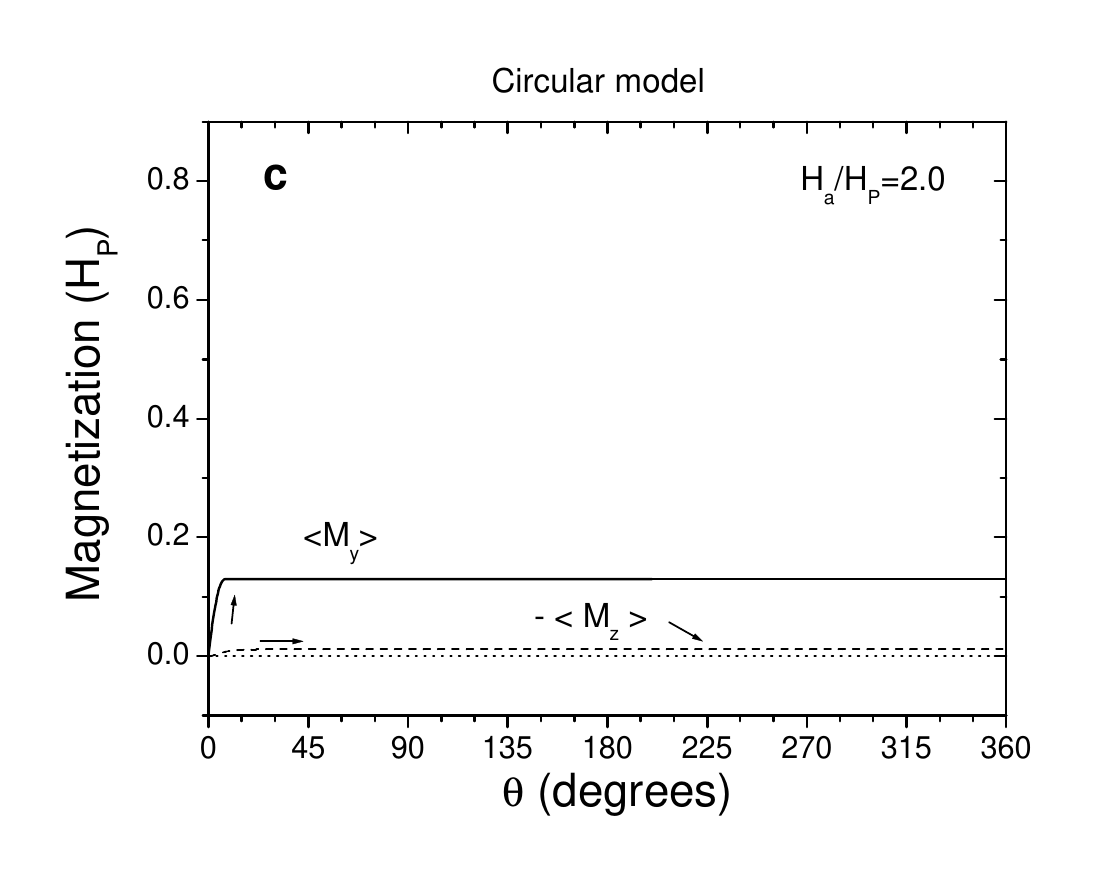}
\captionof{figure}{Curves of the average magnetization components versus the
rotation angle, calculated with Bean's critical-state
model.\label{F3}}
\end{Figure}

\vspace{5mm}
\noindent
Thus, for example, the \ ``oscillations" \ of the magnetization 
components (Fig. \ref{F3},a) have small amplitudes compared with the experimental ones.
Besides, at $H_a=0.8 H_P$ the functions $<M_y>(\theta)$ and $-<M_z>(\theta )$ approximate each
other but at relatively large rotation angles $\theta > 300^{\circ
}$. Finally, when the applied field has an amplitude larger than
$H_p$ (see panel c), the magnetization components are rather small
in magnitude and their initial growth, before the saturation, occurs
in a very small interval of $\theta $ ($<20^{\circ}$).

\subsubsection{Elliptic model}

 The calculations of magnetization components $<M_y>$ and $-<M_z>$ within the elliptic
 flux-line-cutting critical-state model (\ref{ellm}) are shown in Fig. \ref{F4}.
 Here, we used  the same $J_{c\perp}(B)$ as in Eq. (\ref{jcperp}) and
 $J_{c\parallel}(B)$ of the form
\begin{equation}\label{jcpar}
J_{c\parallel}(B)=\frac{J_{c\parallel}(0)}{(1+B/\mu_0H_P)^{n_{\parallel}}}
\end{equation}
with $J_{c\parallel}(0)=1.5J_{c\perp}(0)$ and $n_{\parallel}=1$.
 This choice provides a good agreement between experimental (Fig.~\ref{F2})
 and  theoretical (Fig.~\ref{F4}) curves. Thanks to the use of a second parameter ($J_{c\parallel}$),
 the elliptic model is able to generate the ``oscillations" of the
 magnetization components (Fig. \ref{F4},a) with amplitude close to that observed in
 the experiment (panel (a) in Fig. \ref{F2}). Notice that $<M_y>$ and
 $-<M_z>$ approach each other at $\theta > 150 ^{\circ}$ with
 $H_0=1.05 H_P$ in good concordance with the measurements (see Fig.
 \ref{F2},b, corresponding to $H_0=H_P$). In addition, when $H_0=2.0 H_p$ (panel (c) in Fig. \ref{F4}), the difference between
$<M_y>$ and $-<M_z>$ at $\theta > 45 ^{\circ}$ is as large as in the
experiment (Fig. \ref{F4},c).

\subsubsection{Extended elliptic model}

As was commented in Sec. \ref{Theory}, both elliptic and circular
critical-state models are particular cases of the extended elliptic
one. Therefore, the results presented in Fig. \ref{F3}, predicted by
the circular model, can also be calculated by using the new model
(Eqs. (\ref{eel1}) and (\ref{eel2})) with
$J_{c\perp}=J_{c\parallel}$  as in Eq. (\ref{jcperp}) and
$r=\rho_{\parallel}/\rho_{\perp}=\rho_c/\rho_d$ being equal to one
($r=1$) at $J>J_c$. The condition $r=1$ guarantees that the electric
field ${\bm E}$ and current density ${\bm J}$ be parallel as it is
postulated by Bean's critical-state model (\ref{Bm}). In addition,
graphs in Fig. \ref{F4} (original elliptic model predictions), which
quantitatively reproduce experimental measurements (Fig. \ref{F2}),
are also obtained with the extended elliptic critical-state model
(Eqs. (\ref{eel1}) and (\ref{eel2})) if
$r=J_{c\perp}/J_{c\parallel}$ (i.e.
$\rho_c/\rho_d=J_{c\perp}/J_{c\parallel}$). According to the
parameters $J_{c\perp}(B)$ (\ref{jcperp}) and $J_{c\parallel}(B)$
(\ref{jcpar}), used for calculating magnetization curves in Fig.
\ref{F4}, the ratio $r$ is here smaller than 1 ($r<1$).

It is interesting to study the effect of the parameter $r$,
controlling the relation between the electric field ${\bm E}$


\begin{Figure}
\includegraphics[scale=0.8]{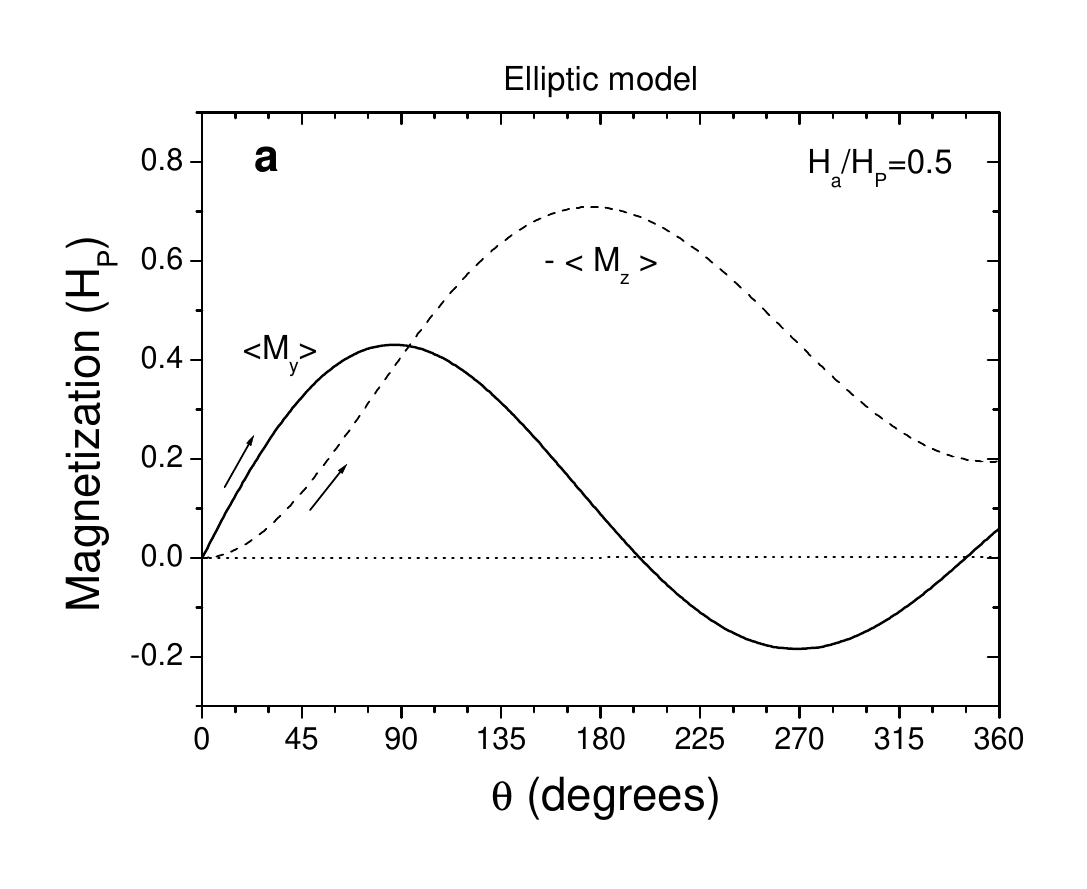}
\vspace{5mm}
\includegraphics[scale=0.8]{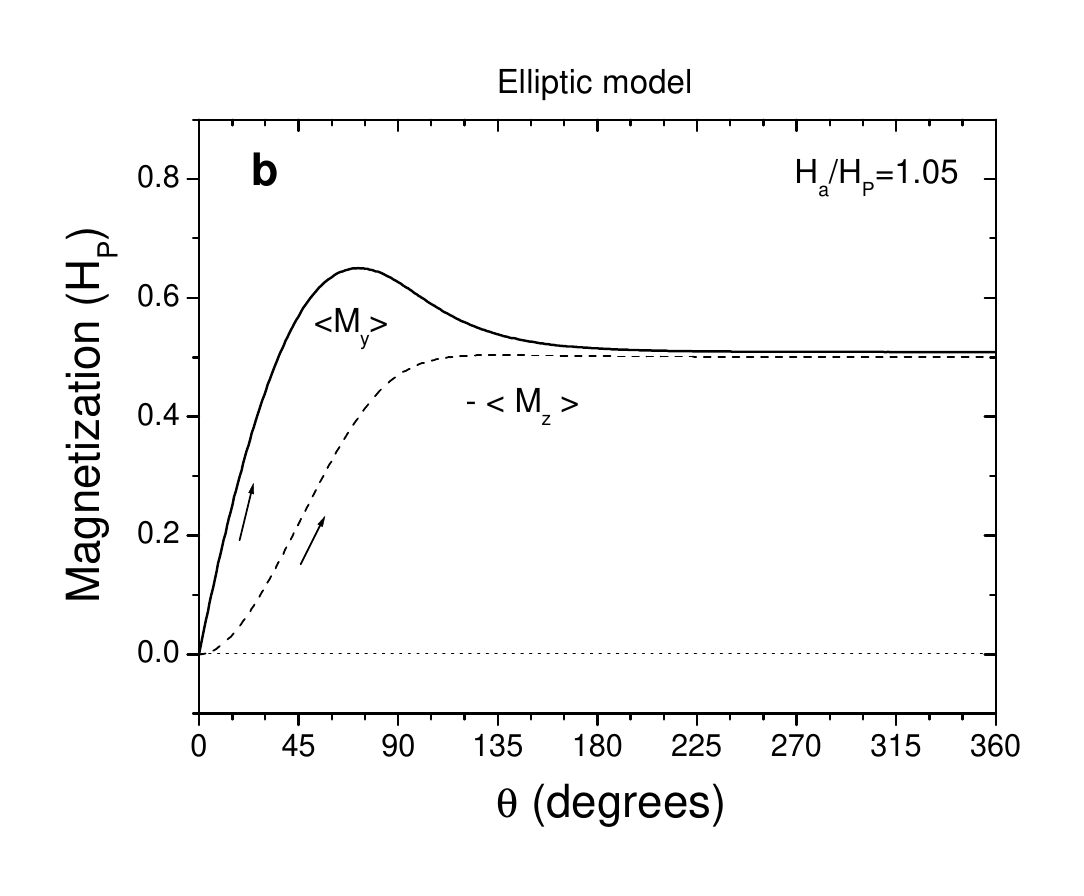}
\vspace{5mm}
\includegraphics[scale=0.8]{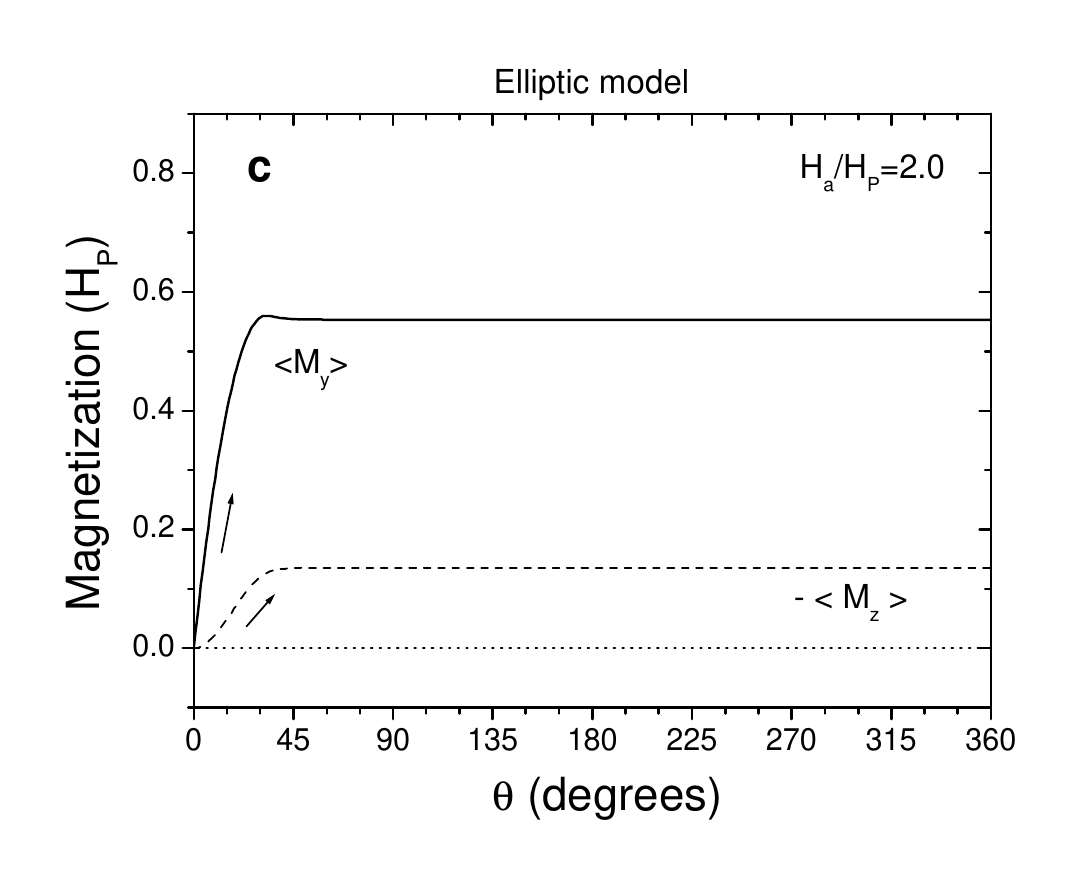}
\captionof{figure}{Curves of the average magnetization components versus the
rotation angle, calculated with the original elliptic critical-state
model.\label{F4}}
\end{Figure}


\begin{Figure}
\includegraphics[scale=0.8]{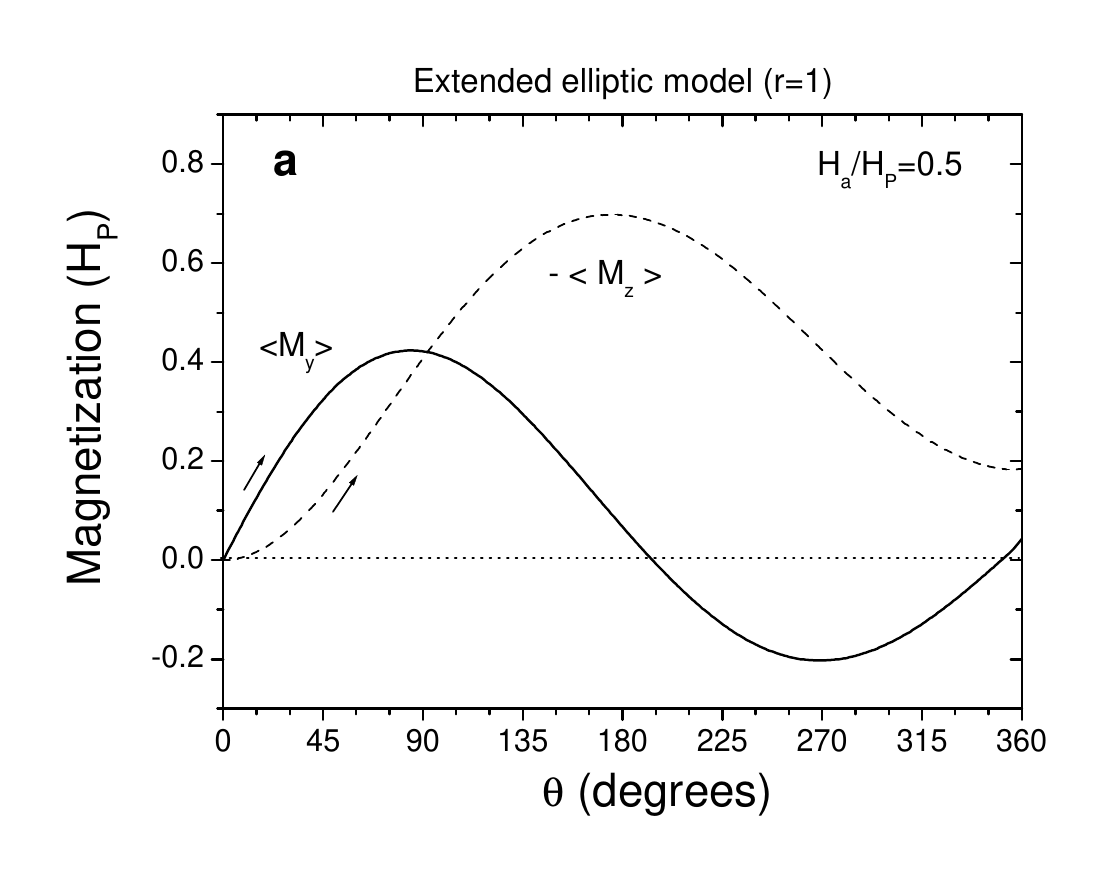}
\vspace{5mm}
\includegraphics[scale=0.8]{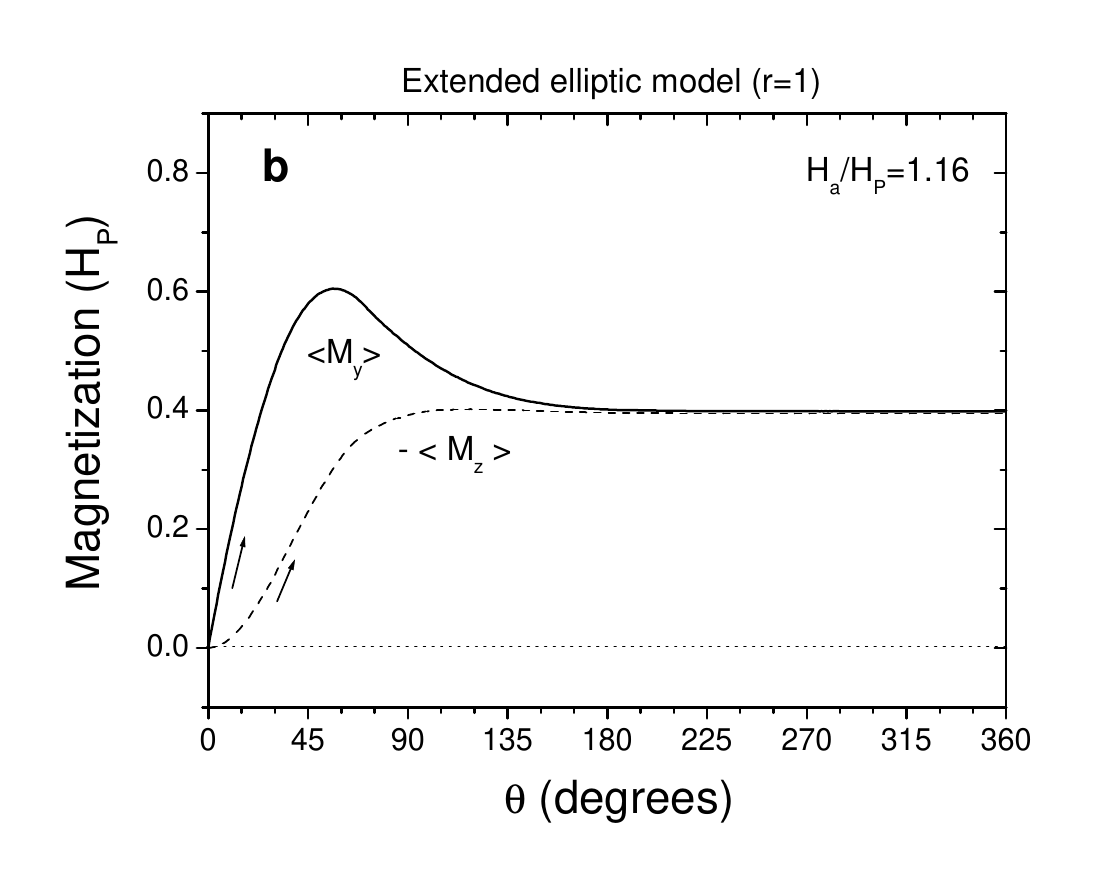}
\vspace{5mm}
\includegraphics[scale=0.8]{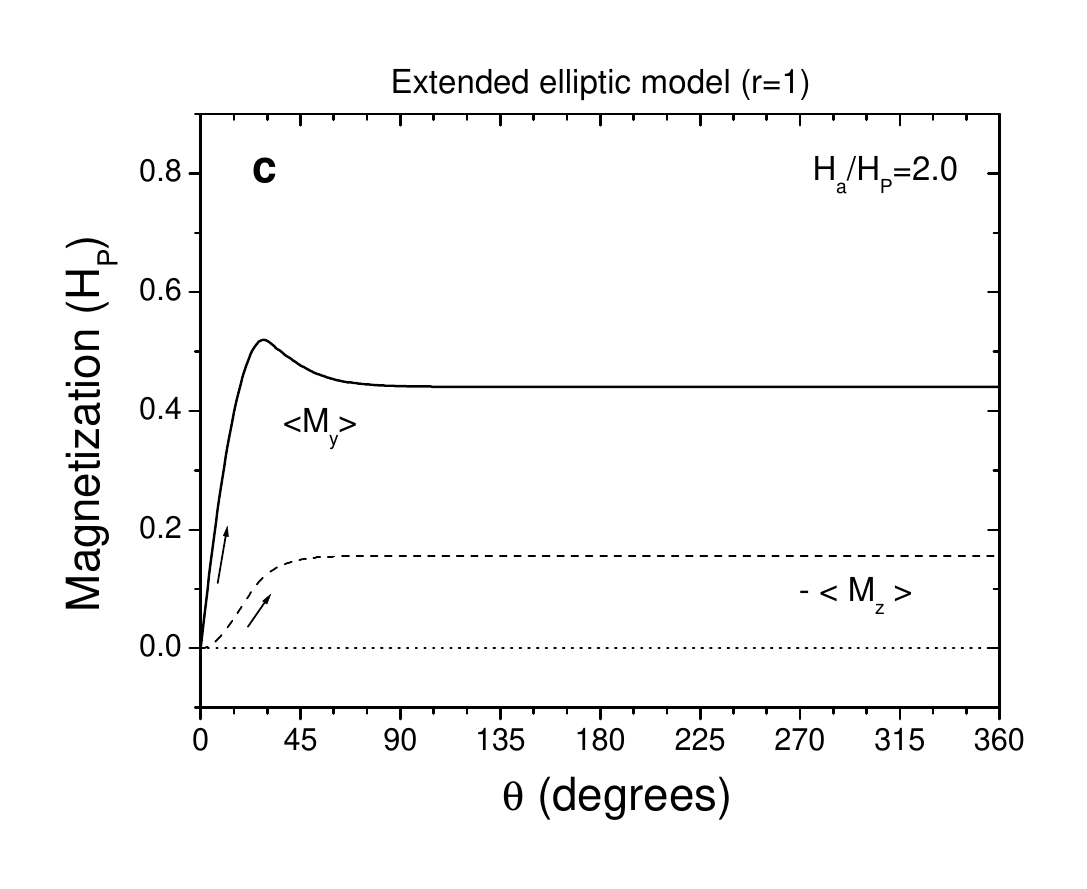}
\captionof{figure}{Curves of the average magnetization components versus the
rotation angle, calculated with the extended elliptic critical-state
model using a ratio $r=1$.\label{F5}}
\end{Figure}

\begin{Figure}
\includegraphics[scale=0.8]{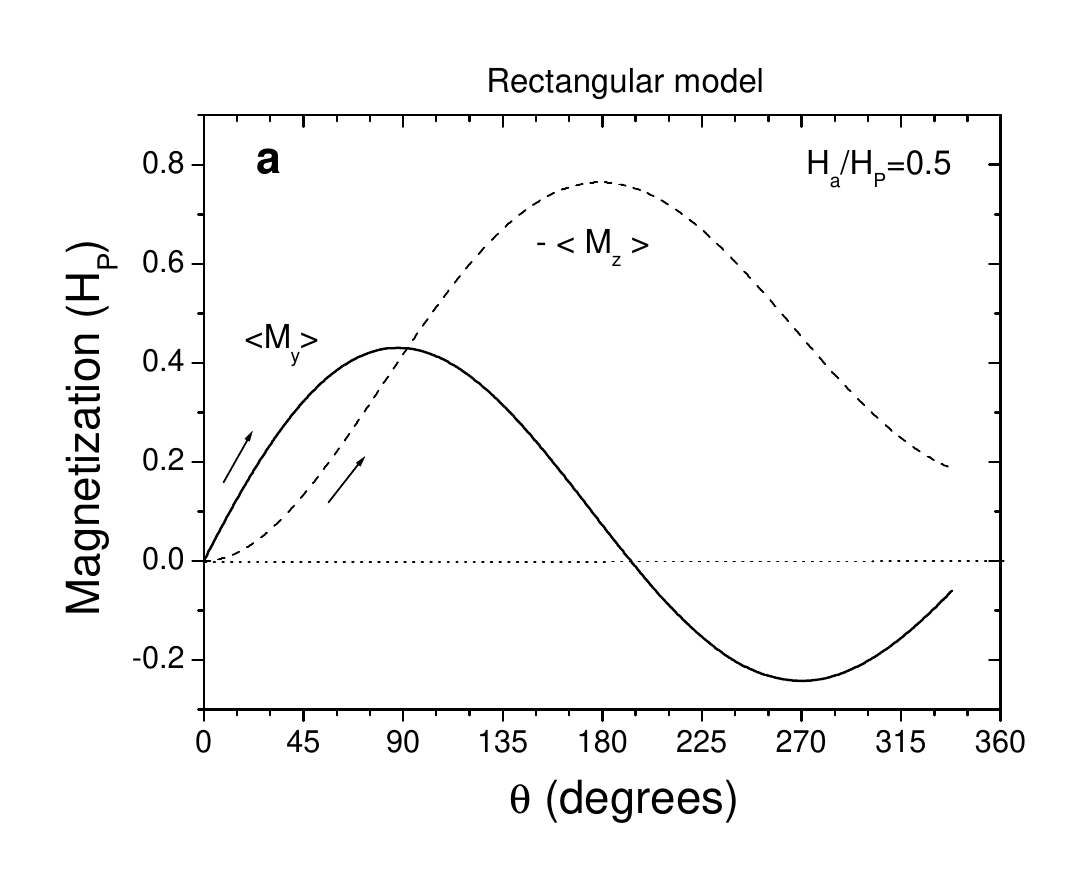}
\vspace{5mm}
\includegraphics[scale=0.8]{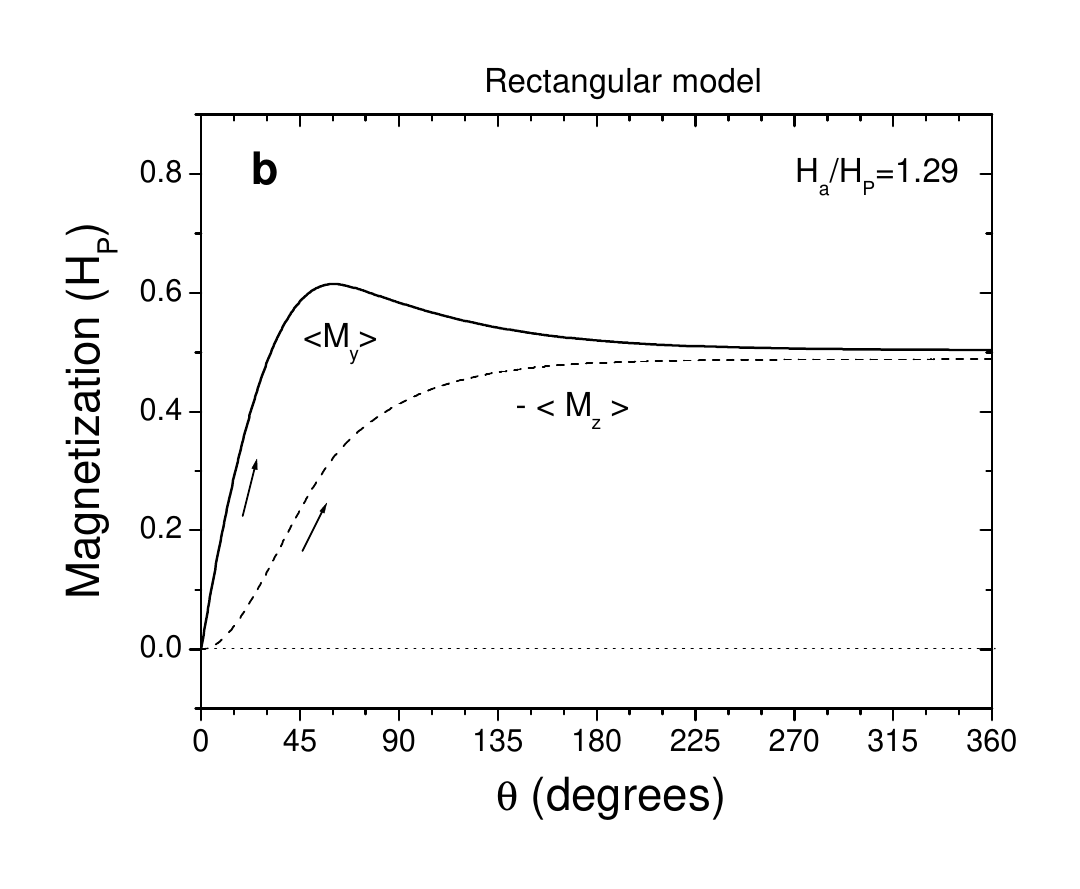}
\vspace{5mm}
\includegraphics[scale=0.8]{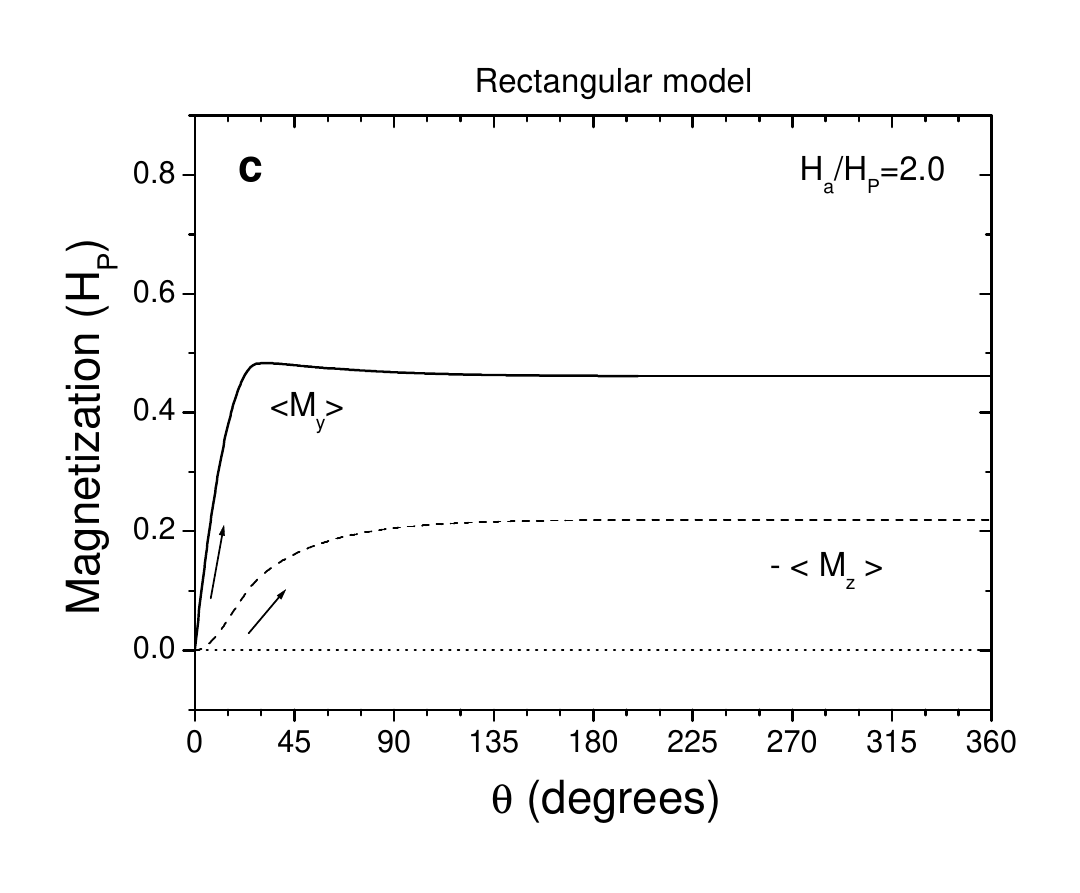}
\captionof{figure}{Curves of the average magnetization components versus the
rotation angle, calculated with the generalized double
critical-state model.\label{F6}}
\end{Figure}

\vspace{10mm}

\noindent
and the current density ${\bm J}$ at $J>J_c$. For this reason, we have
calculated magnetization curves (Fig. \ref{F5}) by applying the
extended elliptic model with the same parameters $J_{c\perp}(B)$ and
$J_{c\parallel}(B)$ as those employed in Fig. \ref{F4}, but with the
parameter $r=\rho_c/\rho_d=1$. In other words, the magnetization curves in Fig. \ref{F5} correspond to an anisotropic critical-state model
with $J_{c\perp}/J_{c\parallel}<1$,
but the parameter $r=1$, indicating that ${\bm E}$ and ${\bm J}$ are parallel when $J>J_c$.
From the comparison of Fig. \ref{F5} with \ref{F4}, we note that
magnetization curves significantly depend upon the parameter $r$
when the applied magnetic field is large enough ($H_a> H_P$ as in
panels (b) and (c)). So, in order the magnetization components,
$<M_y>$ and $-<M_z>$, to have the same value at large angles of
rotation, the applied magnetic field $H_a$ for $r=1$ (Fig.
\ref{F5},b) should be larger than the field used in Fig. \ref{F4},
b. Besides, the value of $<M_y>$ and $-<M_z>$ ($\approx 0.4 H_P$),
at sufficiently large angles $\theta $, turns out to be smaller than
that ($\approx 0.5 H_P$) predicted by the original elliptic model
(Fig. \ref{F4},b). At $H_a=2.0 H_P$, there is also a noticeable
difference between magnetization $y$-components (compare panels (c)
of Figs. \ref{F4} and \ref{F5}).

\subsubsection{Rectangular model}

 For completeness of our study, we have employed the GDCSM (rectangular model), which also uses
 two critical current
densities, namely $J_{c\perp}(B)$ and $J_{c\parallel}(B)$. The
former is determined from the curves of magnetization versus the
applied field, varying along one direction only (Fig. \ref{F1}). In
our case, the magnetic dependence of $J_{c\perp}$ is the same as in
Eq. (\ref{jcperp}). To reproduce the main features of the experiment
(Fig. \ref{F2}), the other parameter is chosen as in Eq.
(\ref{jcpar}), but $J_{c\parallel}(0)=1.32J_{c\perp}(0)$ and
$n_{\parallel}=1.06$ (compare Figs. \ref{F2} and \ref{F6}). Although
these values are different from those used within the elliptic
critical-state model, the parallel critical current density
$J_{c\parallel}$ remains being larger than the perpendicular one
$J_{c\perp}$. It should be noted that the GDCSM predicts the
equality of $<M_y>$ and $-<M_z>$ ($\approx 0.5 H_P$) with an
external field $H_a=1.29H_P> H_P$ at relatively large rotation
angles $\theta > 270^{\circ}$  (see Fig. \ref{F6},b), in contrast to
the experiment where such a behavior occurs from $\theta \approx
150^{\circ}$. Besides, the numerical calculations for $H_a=0.5H_P$
(panel (a) in Fig. \ref{F6}) had to be stopped at $\theta \approx
338^{\circ}$ because the solution further diverged.

\subsection{Magnetic induction profiles}
The fact that the elliptic critical-state model is able to
quantitatively reproduce the experiment, with the use of a parallel
critical current density $J_{c\parallel}(B)$ larger than the
perpendicular one $J_{c\perp}(B)$,  illustrates how flux-line
cutting influences on the magnetic behavior of a rotating
superconductor. To explain the features observed in both
experimental (Fig. \ref{F2}) and theoretical (Fig. \ref{F4})
magnetization curves, we shall analyze the evolution

\end{multicols}

\begin{figure*}
\includegraphics[scale=0.7]{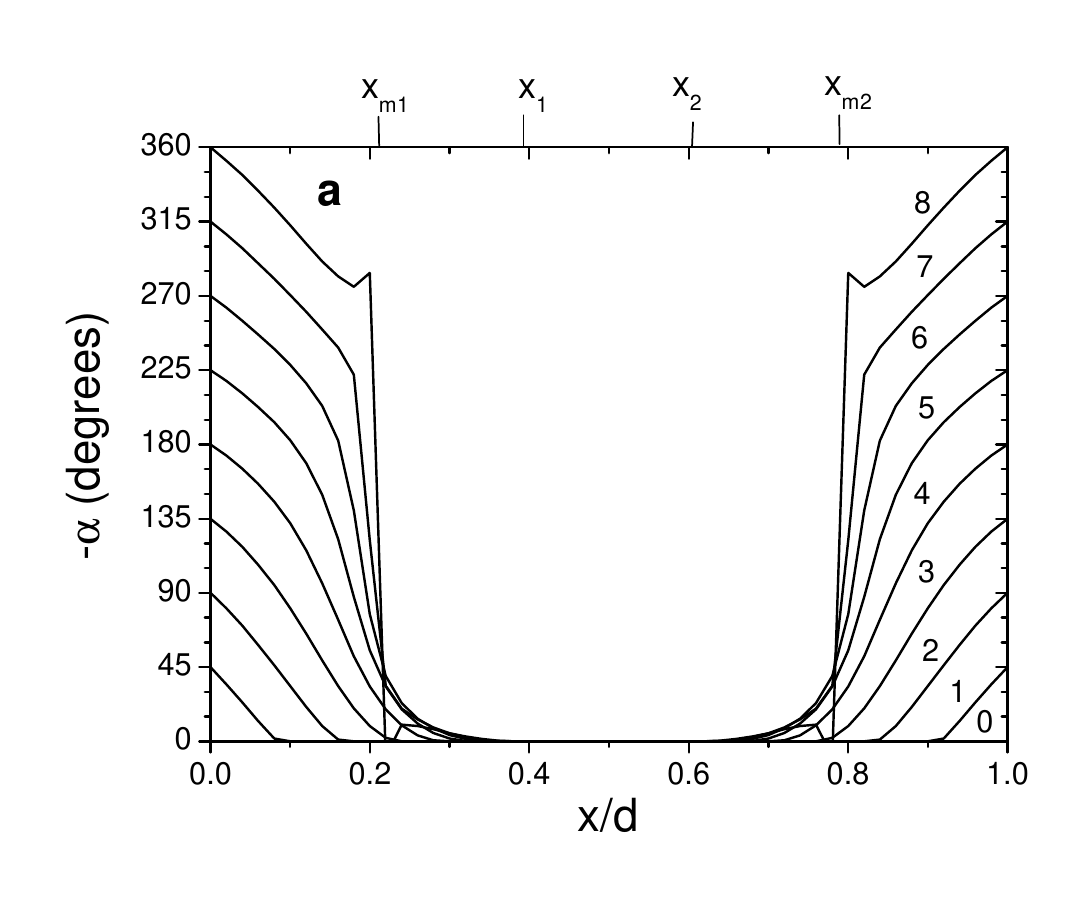}
\includegraphics[scale=0.7]{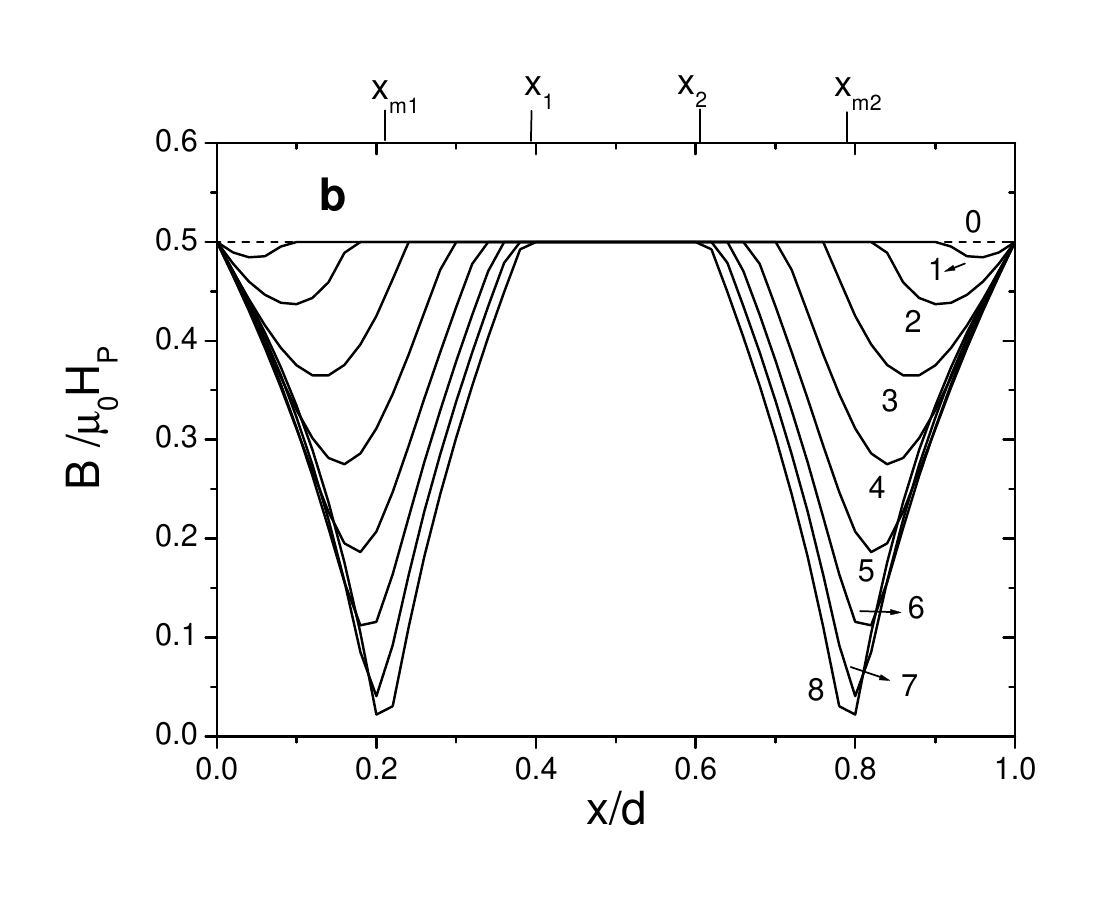}
\includegraphics[scale=0.7]{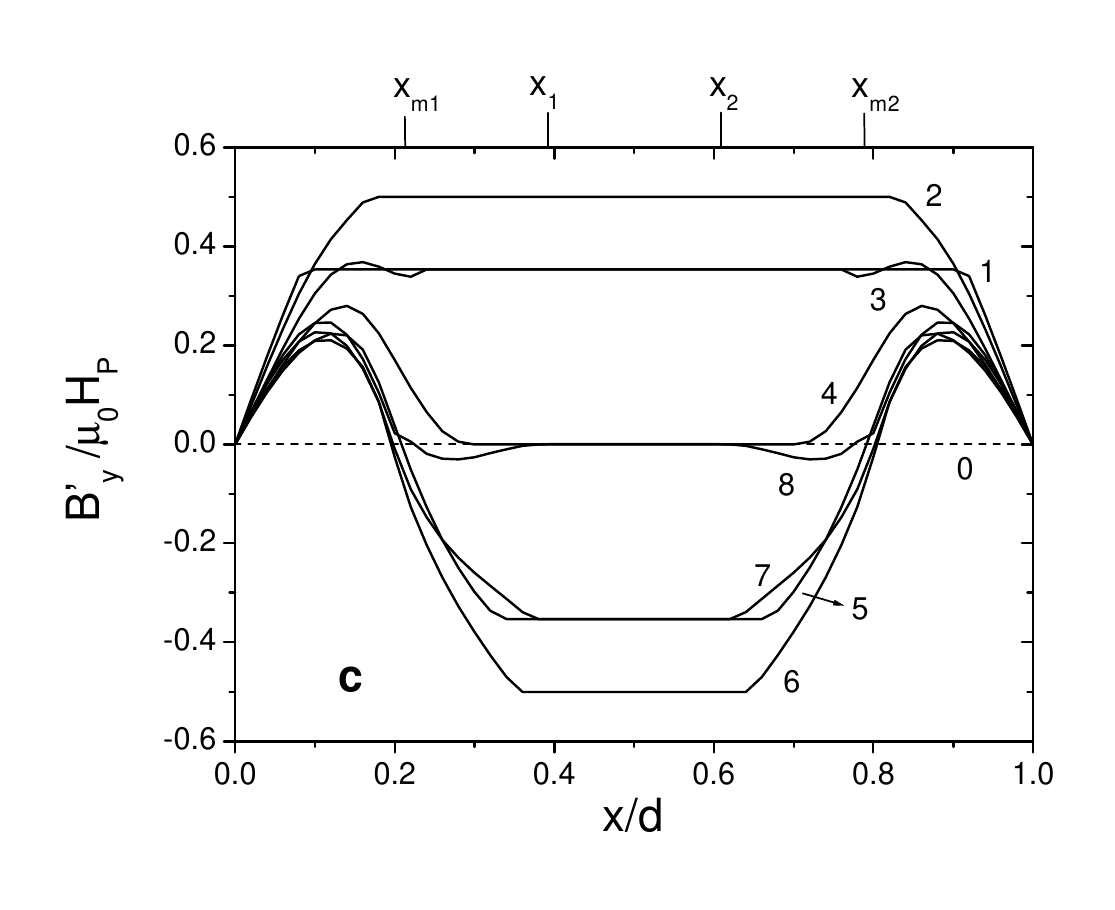}
\includegraphics[scale=0.7]{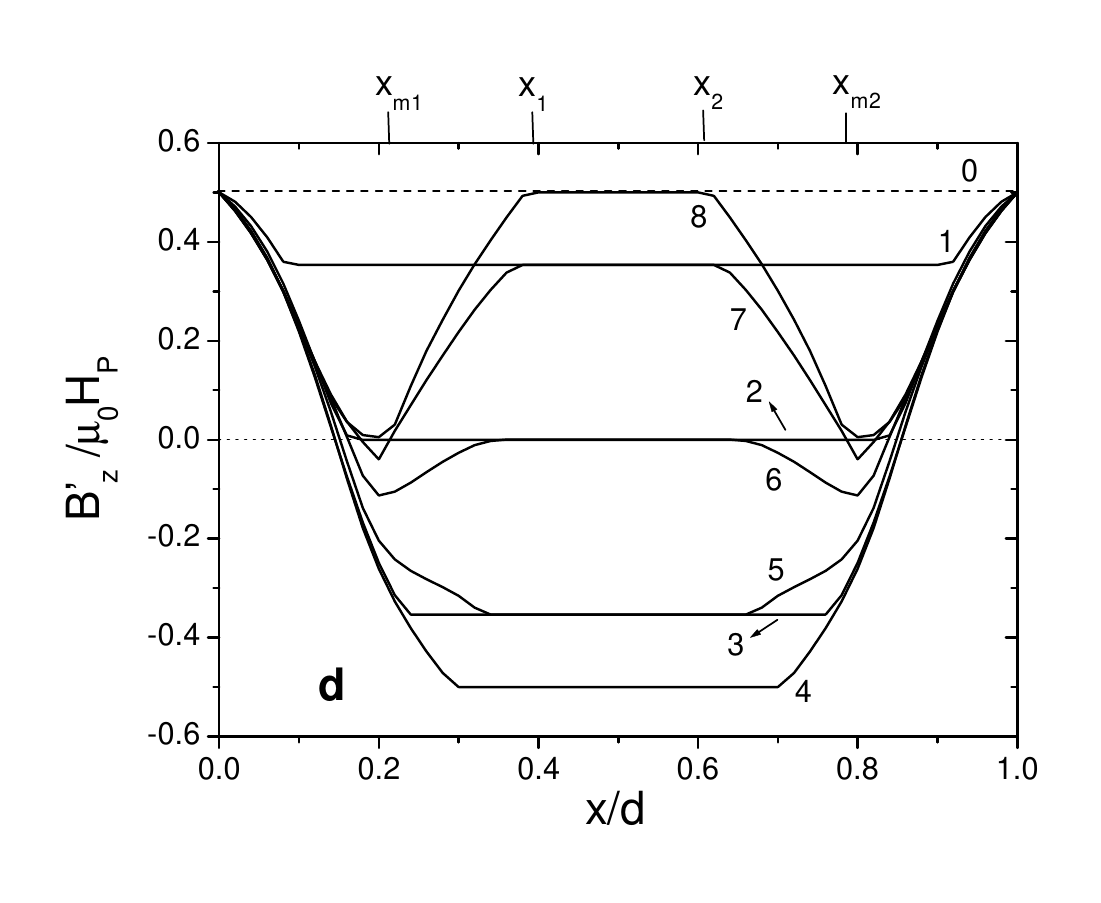}
 \caption{Profiles of the angle $\alpha$ (panel a), magnitude $B$ (panel b) and components
 $B_y'$ (Eq. (\ref{byp}), panel c) and  $B_z'$ (Eq. (\ref{bzp}), panel d) of the magnetic
 induction, calculated with the original elliptic critical-state model at
$H_a=0.5 H_P$.\label{F7}}
\end{figure*}

\begin{figure*}
\includegraphics[scale=0.7]{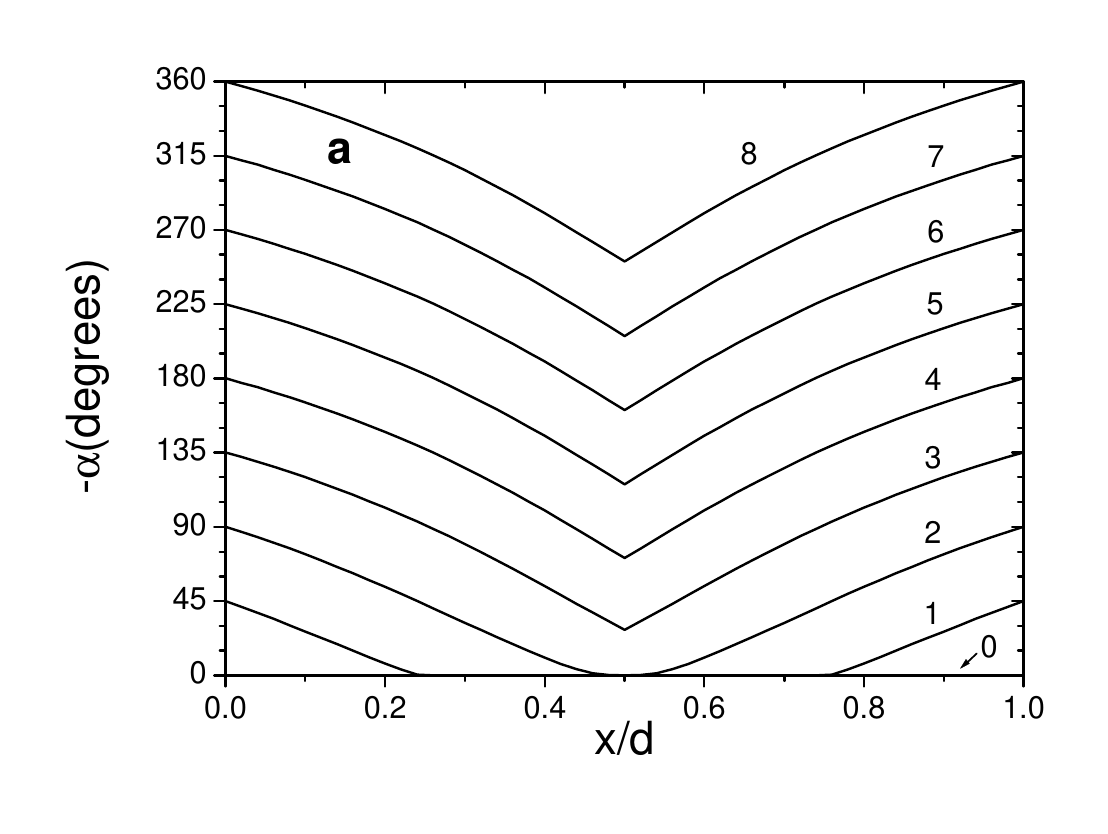}
\includegraphics[scale=0.7]{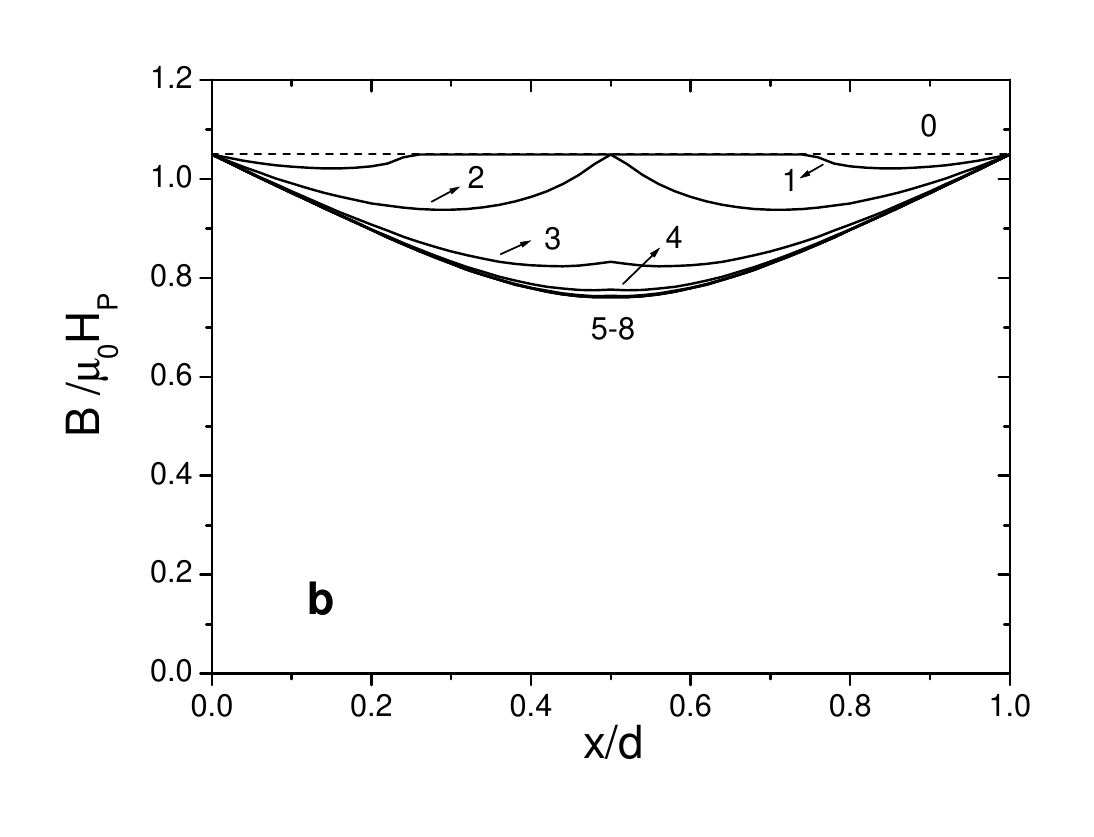}
\includegraphics[scale=0.7]{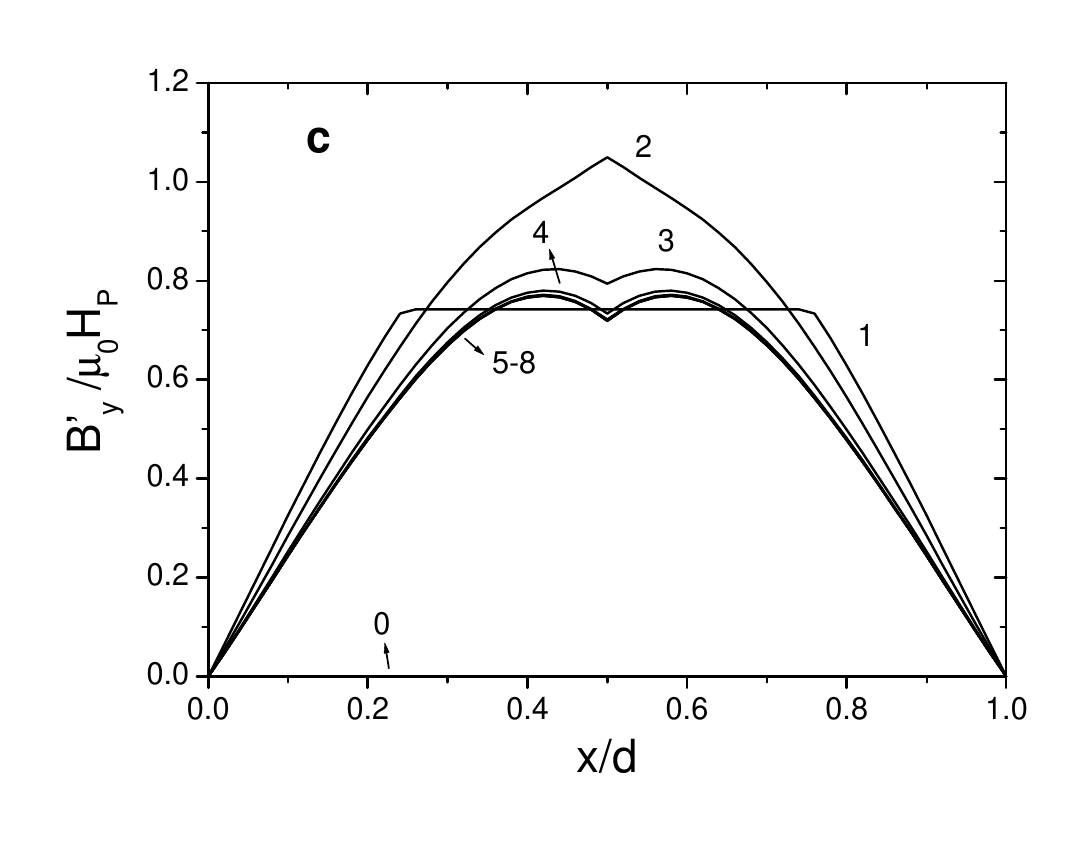}
\includegraphics[scale=0.7]{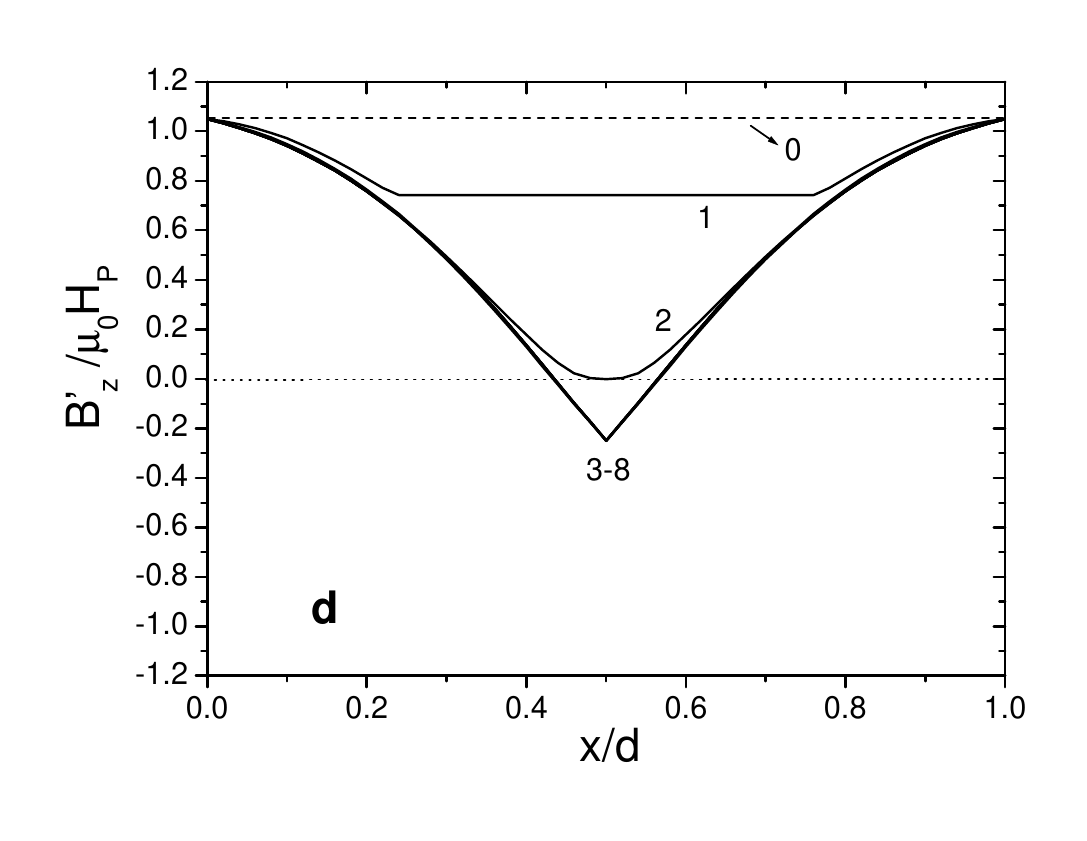}
 \caption{Profiles of the angle $\alpha$ (panel a), magnitude $B$ (panel b) and components
 $B_y'$ (Eq. (\ref{byp}), panel c) and  $B_z'$ (Eq. (\ref{bzp}), panel d) of the magnetic
 induction, calculated with the original elliptic critical-state model at $H_a=1.05 H_P$.\label{F8}}
\end{figure*}

\begin{figure*}
\includegraphics[scale=0.7]{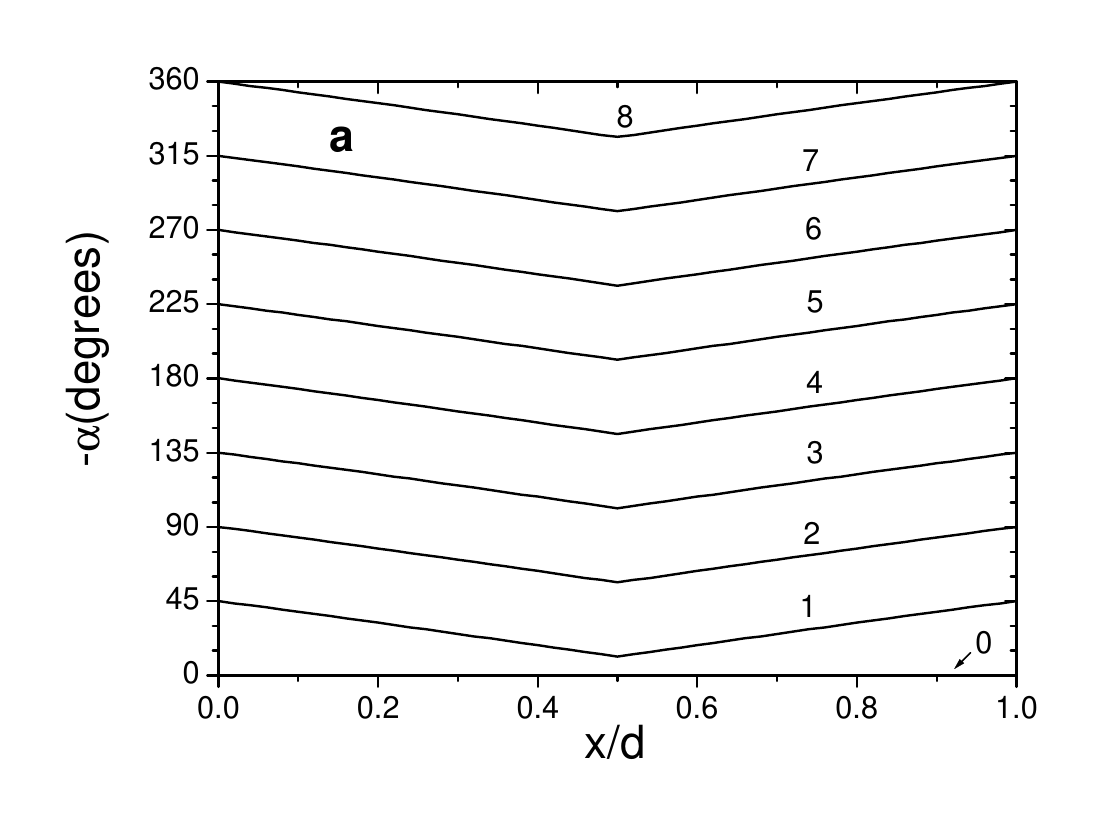}
\includegraphics[scale=0.7]{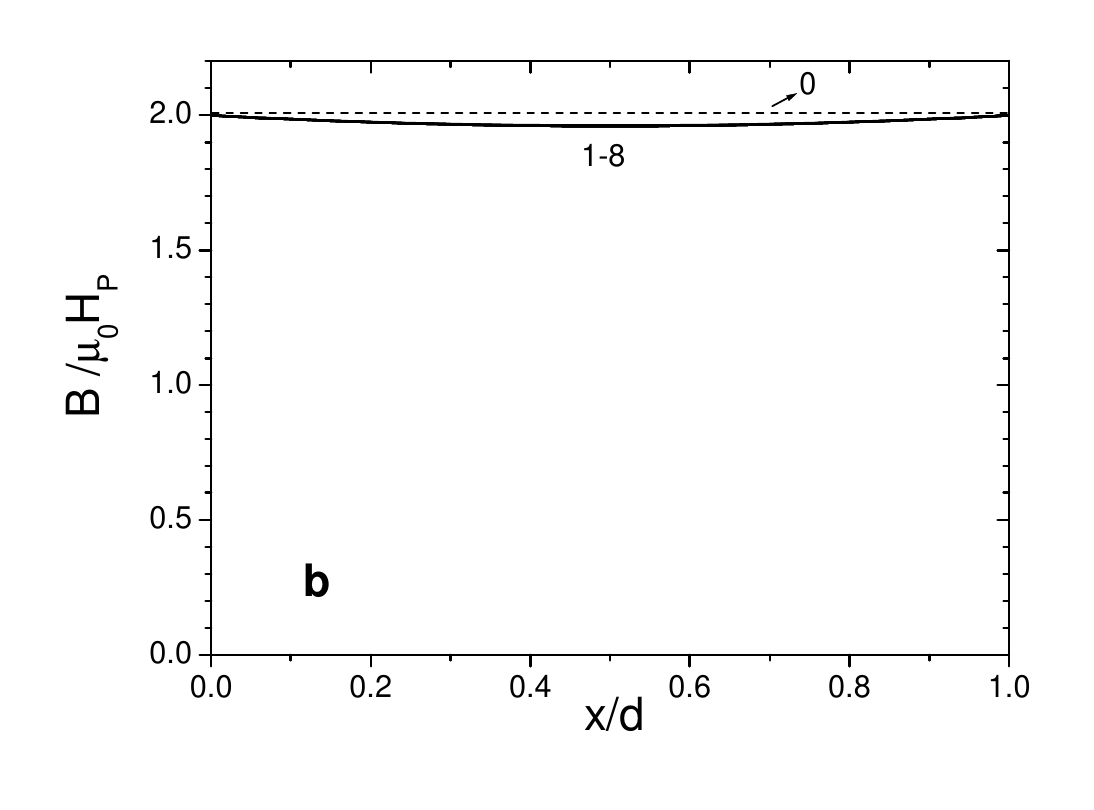}
\includegraphics[scale=0.7]{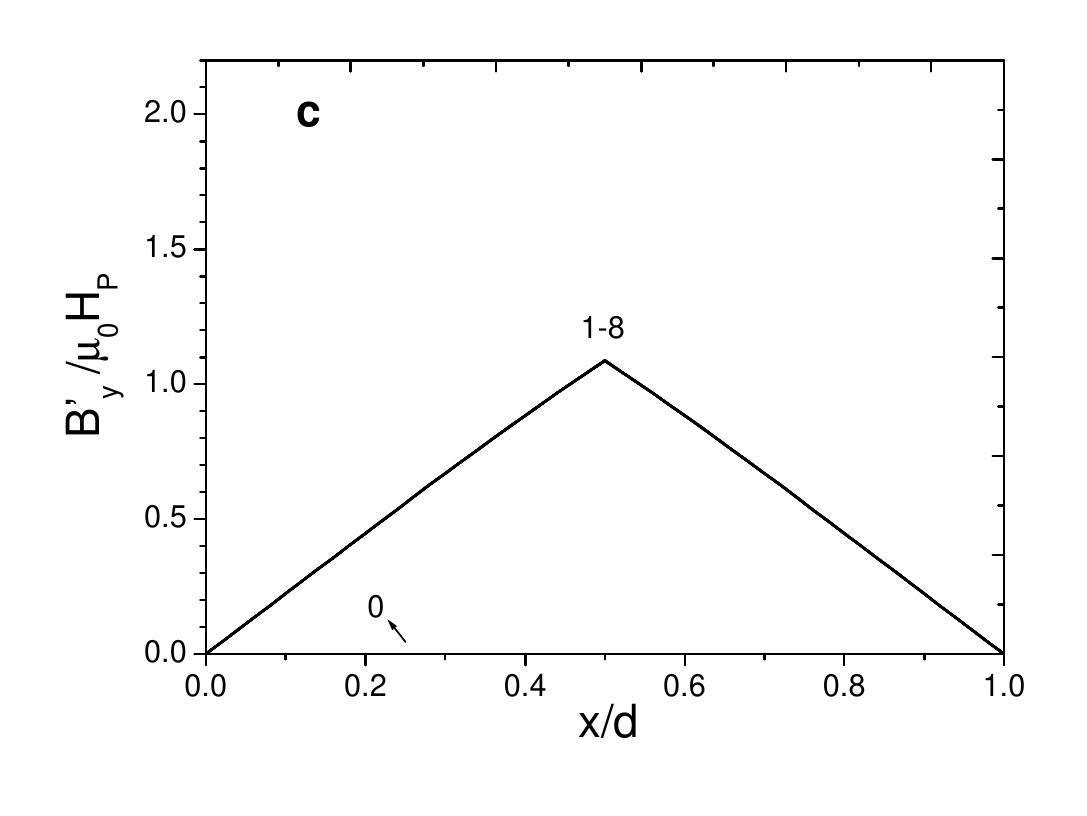}
\includegraphics[scale=0.7]{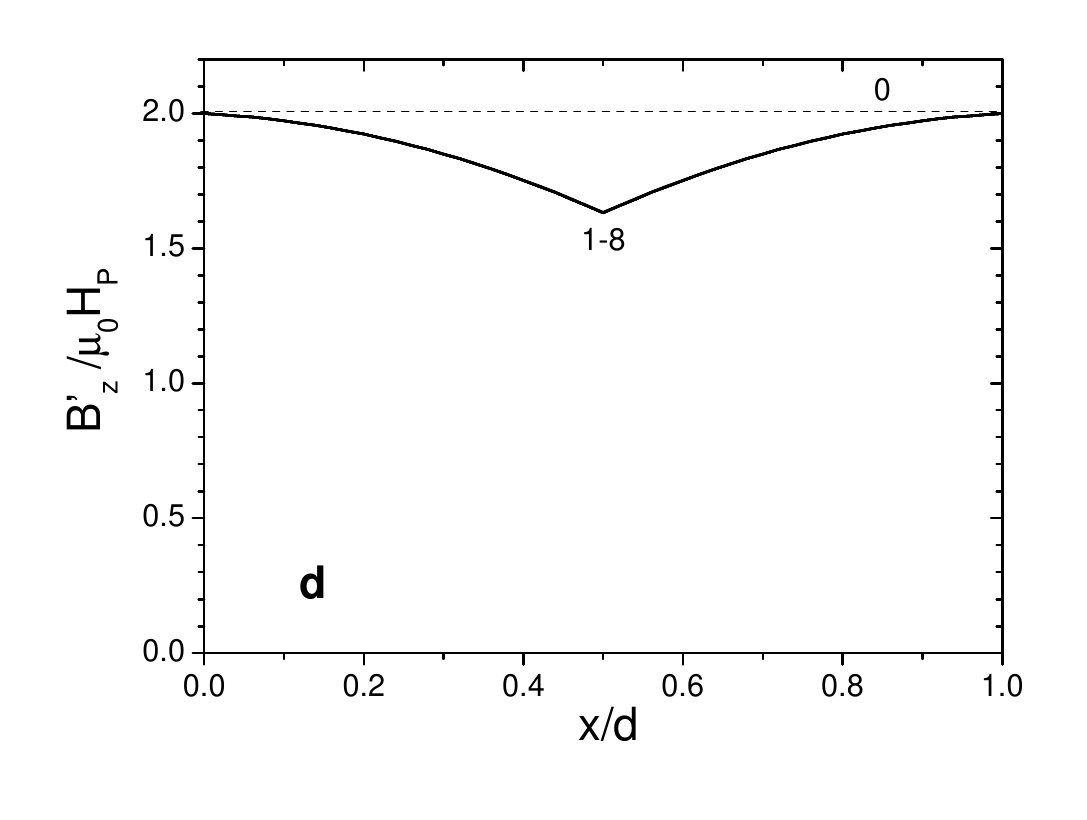}
 \caption{Profiles of the angle $\alpha$ (panel a), magnitude $B$ (panel b) and components
 $B_y'$ (Eq. (\ref{byp}), panel c) and  $B_z'$ (Eq. (\ref{bzp}), panel d) of the magnetic
 induction, calculated with the original elliptic critical-state model at
$H_a=2.0 H_P$.\label{F9}}
\end{figure*}

\begin{multicols}{2}
\noindent
of the profiles
for the magnitude of the magnetic induction $B(x)$, the tilt angle
$\alpha(x)$, and the components $B_y'(x)$ (\ref{byp}) and
$B_z'(x)$(\ref{bzp}), calculated within the original elliptic
flux-line-cutting critical-state model (Figs. \ref{F7}-\ref{F9}).

The calculated profiles of the magnetic induction in the case when
the external magnetic field $H_a$ has a magnitude smaller than  the
penetration field $H_p$ ($H_a=0.5 H_p$) are shown in Fig. \ref{F7}.
As the angle of rotation is increased, two $U$-shaped minima in the
$B(x)$  profile (panel b) appear because of the flux consumption
(decrement of $B$)  which results from flux-line cutting
\cite{Cle82}. The absolute value of the tilt angle $\alpha $
increases with $\theta$ in the near-surface intervals $0\leq x <
x_{m1}$ and $x_{m2}< x\leq d$. However, in the intervals $x_{m1} <
x< x_1$ and $x_2< x < x_{m2}$, where there is flux consumption, the
angle $\alpha $ is slightly modified. In the central interval,
$x_1<x<x_2$, neither $B$ or $\alpha$ is altered. When $\theta
\approx 360 ^{\circ}$, the minimum values of $B$ inside the
superconducting disk tend to zero and, as follows from
Eq.~(\ref{ME2}), the magnitude of the derivative $\partial
\alpha/\partial x$ considerably increases at such points. Besides,
at $x=x_{m1}$ and $x=x_{m2}$ with $B(x_{m1})=B(x_{m2})\approx 0$,
the accuracy of our calculations is low and, therefore, the values
$-\alpha (x_{m1})$ and $-\alpha (x_{m2})$ turned out to be
apparently higher than they should be (see curve 8 for $\theta
=360^{\circ}$ in Fig. \ref{F7},a). The component $B'_z$ of the
magnetic induction, parallel to the applied magnetic field ${\bf
H}_a$, decreases near sample surfaces because of the flux
consumption (Fig.~\ref{F7},d). Nevertheless, the most important
change occurs in the central part of the sample (in $x_1<x<x_2$)
because of the sample rotation. So, at $\theta = 180^{\circ}$ (curve
4) the component $B'_z$ varies from $B'_z=\mu_0 H_a$ at the surfaces
$x=0$ and $x=d$ to the opposite value $B'_z=-\mu_0 H_a$ in the
central region of the sample. When an entire cycle is finished,
$B'_z$ again takes the value $B'_z=\mu_0 H_a$ in the middle of the
disk (curve 8). This cyclic behavior of $B'_z$ is responsible for
the ``oscillations" of the magnetization component $<M_z>(\theta )$
(panels (a) in Figs.~\ref{F2} and \ref{F4}), being negative for any
value of the angle of rotation $\theta
>0$ because $B'_z<\mu_0H_a$ near surfaces, i.e. in the intervals $0<x<x_1$
and $x_2<x<d$. The component  $B'_y$ also oscillates in the middle
of the sample as $\theta$ is increased

\end{multicols}

\begin{multicols}{2}
\noindent
(Fig. \ref{F7},c). Such a
behavior of $B'_y$ makes the magnetization $y$-component $<M_y>$
oscillate with $\theta $ (Figs. \ref{F2},a and \ref{F4},a). As it is
seen in Fig. \ref{F7}c, there is an increment of $B'_y$ in the
near-surface regions, producing a small positive value for $<M_y>$
(\ref{My}) after a complete cycle, i.e. at $\theta = 360^{\circ}$
(see Figs. \ref{F2},a and \ref{F4},a).

Fig. \ref{F8} exhibits profiles calculated within the elliptic
critical-state model for $H_a = 1.05 H_p$. Due to the decrease of
the critical current densities  $J_{c\perp}$ (\ref{jcperp}) and
$J_{c\parallel}$ (\ref{jcpar}) with the magnitude $B$ of the
magnetic induction, the slopes of the critical profiles for $B(x)$
and $\alpha (x)$ near surfaces are smaller than the slopes observed
in the corresponding profiles of  Fig. \ref{F7}. Therefore, the
central region with unaltered $B$ and $\alpha $ (see curves 1 in
panels (a) and (b) of Fig. \ref{F8}) rapidly disappears as the
rotation angle $\theta $ is increased (see curves 2 therein). Also,
the $U$-shaped minima of $B(x)$ coalesce forming a unique minimum at
the center of the disk. The resulting critical profile $B(x)$ does
not further change despite the fact that the disk continues rotating
(see curves 5-8 in panel (b)). In this case, $B'_z(x)$ initially
decreases (curves 1-2 in Fig. \ref{F8},d) inside the sample as
$\theta $ varies until it reaches the critical profile (curves 3-8).
Hence, the dependence $<M_z> (\theta )$ has a monotonic behavior at
$\theta > 120^{\circ} $ (see panels (b) in Figs.~\ref{F2} and
\ref{F4}). On the other hand, $B'_y(x)$ increases so that a huge
maximum in the dependence $<M_y>(\theta ) $ (Figs.~\ref{F2},b and
\ref{F4},b) appears at $\theta \approx 70^{\circ}$. At large
rotation angles ($\theta > 180 ^{\circ}$), the profile $B'_y(x)$
becomes stationary and $<M_y>(\theta )$ is, practically, a constant
function, having a value close to $-<M_z>$. So, the magnitude of the
magnetization, $|<{\bf M}>|$, is independent of $\theta $ when the
rotation angle is sufficiently large.

The profiles for the case when the external magnetic field is large
enough, in comparison with the penetration field $H_P$ (as in
Fig.~\ref{F9}), have an evolution similar to that presented in Fig.
\ref{F8}. However, the central regions of unaltered magnetic
induction rapidly disappear as $\theta $ is increased, in comparison
with the results of Fig. \ref{F8}. This fact is due to noticeable
reduction of the critical current densities $J_{c\perp}$ and
$J_{c\parallel}$ with $B$.

\section{Conclusion}

We have applied the circular, elliptic, extended-elliptic, and
rectangular critical-state models to study the magnetic behavior of
irreversible type-II superconductors

\end{multicols}

\begin{multicols}{2}
\noindent
in a parallel rotating magnetic
field. The numerical method employed here is based on the
substitution of the vertical law, relating the electric field ${\bm
E}$ and the current density ${\bm J}$, for a nonlinear material
equation having effective flux-cutting and flux-flow resistivities
in the dissipative region. The substitution is justified when the
applied magnetic field $H_a$ slowly varies either in magnitude or
direction, inducing  electric fields of sufficiently small magnitude
inside the superconductor. Within the elliptic (circular)
critical-state model such resistivities are not independent of each
other and have a ratio $r=\rho_{\parallel}/\rho_{\perp}$ equal to
$J_{c\perp}/J_{c\parallel}$ (=1 for the circular model) at $J$ just
above its critical value $J_c$.  On the other hand, within the
extended elliptic critical-state model the ratio $r$ is an
independent parameter to be determined. The rectangular
critical-state model also uses two independent resistivities,
$\rho_{\parallel}$ and $\rho_{\perp}$. However, unlike the other
critical-state models, the GDCSM  assumes that flux cutting and flux
depinning do not affect each other.

    The comparison of the predictions of the mentioned critical-state models
    with experimental measurements of magnetization for  a rotating PbBi  disk
    in a fixed magnetic field \cite{Sek89} shows that the original critical-state model
    can reproduce the main features of the magnetization curves. The
    circular and rectangular critical-state models only achieve a
    qualitative description of the experiment. The extended elliptic
    model, being more general than the original elliptic one, has allowed us to
    study the effect of the relation between ${\bm E}$ and ${\bm J}$ in the dissipative
    region. However, additional theoretical and experimental studies are
    needed to elucidate on the effects associated with both flux-cutting and flux-flow
    resistivities.

\subsection*{Acknowledgements}

This work was partially supported by Consejo Nacional de Ciencia y
Tecnolog\'{\i}a (CONACYT, Mexico).

\end{multicols}


\begin{thebibliography}{99}

\bibitem{Fis96} L.~M. Fisher, A.~V. Kalinov, I.~F. Voloshin, I.~V.
Baltaga, K.~V. Il'enko, V.~A. Yampol'skii, Solid State Commun. {\bf
97}, 833 (1996).

\bibitem{Bea62} C.~P. Bean, Phys. Rev. Lett. {\bf 8}, 250 (1962).

\bibitem{Bea70} C.~P. Bean, J. Appl. Phys. {\bf 41}, 2482 (1970).

\bibitem{Cle82} J.~R. Clem, Phys. Rev. B {\bf 26}, 2463 (1982).

\bibitem{Cle84} J.~R. Clem and A. P\'erez-Gonz\'alez, Phys. Rev. B {\bf 30}, 5041
(1984).

\bibitem{Per85a} A. P\'erez-Gonz\'alez and J.~R. Clem, Phys. Rev. B {\bf
31}, 7048 (1985).

\bibitem{Per85b} A. P\'erez-Gonz\'alez and J.~R. Clem, Phys. Rev. B {\bf 32}, 2909 (1985).

\bibitem{Per85c} A. P\'erez-Gonz\'alez and J.~R. Clem, J. Appl.
Phys. {\bf 58}, 4326 (1985).

\bibitem{Wal72} D.~G. Walmsley, J. Phys. F {\bf 2}, 510
(1972).

\bibitem{Cam72} A.~M. Campbell and J.~E. Evetts, Adv. Phys.
{\bf 21}, 199 (1972).

\bibitem{Cav82} J.~R. Cave and M.~A.~R. LeBlanc, J. Appl. Phys. {\bf 53}, 1631 (1982).

\bibitem{Boy77} R. Boyer and M.~A.~R. LeBlanc, Solid State Commun. {\bf 24}, 261 (1977).

\bibitem{Boy80} R. Boyer, G. Fillion, and M.~A.~R. LeBlanc, J. Appl. Phys. {\bf 51}, 1692 (1980).

\bibitem{LeB84} M. A. R. LeBlanc and J. P. Lorrain, J. Appl. Phys.
{\bf 55}, 4035 (1984).

\bibitem{Per97} F. P\'erez-Rodr\'{\i}guez, A. P\'erez-Gonz\'alez, J.~R. Clem, G. Gandolfini, and M.~A.~R. LeBlanc, Phys. Rev. B {\bf 56}, 3473 (1997).

\bibitem{Sil98} A. Silva-Castillo, R.~A. Brito-Orta, A. P\'erez-Gonz\'alez, and
    F. P\'erez-Rodr\'{\i}guez, Physica C {\bf 296}, 75 (1998).

\bibitem{Fis00} L.~M. Fisher, K.~V. Il'enko, A.~V. Kalinov, M.~A.~R.
LeBlanc, F. P\'erez-Rodr\'{\i}guez, S.~E. Savel'ev, I.~F. Voloshin,
V.~A. Yampol'skii, Phys. Rev. B {\bf 61}, 15382 (2000).

\bibitem{Vol10} I.~F. Voloshin, L.~M. Fisher, V.~A. Yampol'skii, Low Temp. Phys. {\bf
36}, 39 (2010).

\bibitem{Rom03c} C. Romero-Salazar and F. P\'erez-Rodr\'{\i}guez, Appl. Phys. Lett. {\bf 83}, 5256 (2003).

\bibitem{Vol01} I.~F. Voloshin, A.~V. Kalinov, L.~M. Fisher, A.~V. Aksenov, and
V.~A. Yampol'skii, JETP {\bf 93}, 1105 (2001).

\bibitem{Rom03b} C. Romero-Salazar and F. P\'erez-Rodr\'{\i}guez,
 Supercond. Sci. Technol. {\bf 16}, 1273 (2003).

\bibitem{Rom04} C. Romero-Salazar and F.
P\'erez-Rodr\'{\i}guez, Physica C {\bf 404}, 317 (2004).

\bibitem{Fis97a} L.~M. Fisher, A.~V. Kalinov, S.~E. Savelev, I.~F. Voloshin,
V.~A. Yampol'skii, M.~A.~R. LeBlanc, and S. Hirscher, Physica C {\bf
278}, 169 (1997).

\bibitem{Fis97b} L.~M. Fisher, A.~V. Kalinov, S.~E. Savelev, I.~F. Voloshin,
and V.~A. Yampol'skii, Solid State Commun. {\bf 103}, 313 (1997).

 \bibitem{Rom05} C. Romero-Salazar, L.~D. Valenzuela-Alacio, A.~F. Carballo-S\'anchez, and F. P\'erez-Rodr\'{\i}guez,  J. Low Temp. Phys. {\bf 139}, 273
(2005).

\bibitem{Lor79} J.~P. Lorrain, M.~A.~R. LeBlanc, and A. Lachaine, Can. J. Phys.
{\bf 57}, 1458 (1979).

\bibitem{Rom08} C. Romero-Salazar and O.~A. Hern\'andez-Flores, J. Appl. Phys. {\bf 103}, 093907 (2008).

\bibitem{Cle11b} J.~R. Clem, M. Weigand, J.~H. Durrell, and A.~M.
Campbell, arXiv:1103.1393v1 [cond-mat.supr-con] 7 Mar 2011.

\bibitem{Bra07} E.~H. Brandt and G.~P. Mikitik, Phys. Rev. B {\bf 76}, 064526 (2007).

\bibitem{Mik10} G.~P. Mikitik, Low Temp. Phys. {\bf 36}, 13 (2010).

\bibitem{Cle11a} J.~R. Clem, arXiv:1102.3678v1 [cond-mat.supr-con] 17 Feb
2011.

\bibitem{Bad09} A. Bad\'{\i}a-Maj\'os, C. L\'opez, and H.~S. Ruiz, Phys. Rev. B {\bf 80}, 144509 (2009).

\bibitem{Sek89} J. Sekerka, M.Sc. thesis ``Flux cutting in semi-reversible and irreversible type II superconductors", University of Ottawa, 1989.

\bibitem{Rom03a} C. Romero-Salazar and F. P\'erez-Rodr\'{\i}guez,
 J. Non-Cryst. Sol. {\bf 329}, 159 (2003).

\end{thebibliography}
\end{document}